\documentclass{sig-alternate}
\usepackage{times}
\usepackage{amsfonts}
\usepackage{graphicx}

\usepackage{url}
\usepackage{epsfig}
\usepackage{graphicx}
\usepackage{xspace}
\usepackage{subfigure}
\usepackage{float}
\usepackage{amsmath}
\usepackage{amssymb}
\usepackage{latexsym}
\usepackage{algorithm}
\usepackage[noend]{algorithmic}
\usepackage{mathrsfs}
\usepackage{wasysym}
\usepackage{color}

\long\def\comment#1{}

\def\thepage{\arabic{page}}
\title{Random-walk domination in large graphs: problem definitions and fast solutions}
\author{
\alignauthor Rong-Hua Li, Jeffrey Xu Yu, Xin Huang, Hong Cheng \\
       \affaddr{The Chinese University of Hong Kong, Hong Kong, China} \\
       \email{\{rhli, yu, xhuang, hcheng\}@se.cuhk.edu.hk}
}

\newcounter{example}[section]
\renewcommand{\theexample}{\nthesection.\arabic{example}}
\newenvironment{example}{
     \refstepcounter{example}
     {\vspace{1ex} \noindent\bf  Example  \theexample:}}{
     \eop\vspace{1ex}} 

\newcounter{definition}[section]
\renewcommand{\thedefinition}{\nthesection.\arabic{definition}}
\newenvironment{definition}{
     \refstepcounter{definition}
     {\vspace{1ex} \noindent\bf  Definition  \thedefinition:}}

\newcounter{theorem}[section]
\renewcommand{\thetheorem}{\nthesection.\arabic{theorem}}
\newenvironment{theorem}{\begin{em}
        \refstepcounter{theorem}
        {\vspace{1ex} \noindent\bf  Theorem  \thetheorem:}}{
        \end{em}}\vspace{1ex} 

\newcounter{lemma}[section]
\renewcommand{\thelemma}{\nthesection.\arabic{lemma}}
\newenvironment{lemma}{\begin{em}
        \refstepcounter{lemma}
        {\vspace{1ex}\noindent\bf Lemma \thelemma:}}{
        \end{em}}

\newcounter{corollary}[section]
\renewcommand{\thecorollary}{\nthesection.\arabic{lemma}}

\newcounter{remark}[section]
\renewcommand{\theremark}{\thesection.\arabic{remark}}

\newcommand{\nthesection}{\arabic{section}}

\newcommand{\eop}{\hspace*{\fill}\mbox{$\Box$}}

\newcommand{\stitle}[1]{\vspace{1ex} \noindent{\bf #1}}

\begin{document}

\thispagestyle{plain}

\maketitle
\begin{abstract}
We introduce and formulate two types of random-walk domination problems in graphs motivated by a number of applications in practice (e.g., item-placement problem in online social network, Ads-placement problem in advertisement networks, and resource-placement problem in P2P networks). Specifically, given a graph $G$, the goal of the first type of random-walk domination problem is to target $k$ nodes such that the total hitting time of an $L$-length random walk starting from the remaining nodes to the targeted nodes is minimal. The second type of random-walk domination problem is to find $k$ nodes to maximize the expected number of nodes that hit any one targeted node through an $L$-length random walk. We prove that these problems are two special instances of the submodular set function maximization with cardinality constraint problem. To solve them effectively, we propose a dynamic-programming (DP) based greedy algorithm which is with near-optimal performance guarantee. The DP-based greedy algorithm, however, is not very efficient due to the expensive marginal gain evaluation. To further speed up the algorithm, we propose an approximate greedy algorithm with linear time complexity w.r.t.\ the graph size and also with near-optimal performance guarantee. The approximate greedy algorithm is based on a carefully designed random-walk sampling and sample-materialization techniques. Extensive experiments demonstrate the effectiveness, efficiency and scalability of the proposed algorithms.
\end{abstract}

\section{Introduction}
\label{sec:intro}
Given a graph $G=(V, E)$ with $n=|V|$ nodes and $m=|E|$, how can we quickly target $k$ nodes such that the targeted nodes can be easily reached by the remaining nodes through $L$-length random walk where the random-walk moves at most $L$ hops? And how can we rapidly find $k$ nodes so as to maximize the expected number of nodes that hit any one targeted node by the $L$-length random walk? We refer to these two problems as two types of random-walk domination problems, because a node hits any one targeted node can be regarded as that the targeted nodes dominate such a node by an $L$-length random walk. Intuitively, the random-walk domination problems are very hard because there are $C_n^k $ possible solutions and for each solution one should perform $n-k$ calculations to check (or record the hitting time) whether or not a node reaches any one targeted node by the $L$-length random walk. These problems are encountered in many data mining and social network analysis applications. Some of them are discussed as follows.

\subsection{Motivation}
\stitle{Item-placement problem in online social networks}: Recently, social networking services are becoming an important media for users to search for information online \cite{07icwsmsocialbrowsing, 07socialbrowsing, 10icwsmsocialsearch, 10vldbsocialsearch, 12cacmsocialsearch}. In many online social networks, users find information primarily rely on a social process called social browsing \cite{07icwsmsocialbrowsing, 07socialbrowsing}. In particular, social browsing depicts a process that the users in a social network find information along their social ties \cite{07icwsmsocialbrowsing, 07socialbrowsing}. For example, in an online photo-sharing website Flickr (\url{http://www.flickr.com/}), a user can view his friends' photos via visiting their home-page. Once the user arrives at one of his friends' home-page, then he is also able to apply the same way to browse the photos created by his friend's friends. Clearly, the next home-page that a user visits only depends on the current home-page that the user stays. Therefore, a user's social browsing process can be regarded as a random-walk process on the social network. Furthermore, users typically has an \emph{implicit} time limit to browse the others' home-pages because users cannot browse infinite number of home-pages. As a result, we can model the social browsing process as an $L$-length random walk by assuming that each user visits at most $L$ home-pages in a social browsing process.

Based on the social browsing process, two interesting questions are: (1) how to place items (e.g., news, photos, videos, and applications) on a small fraction of users in a social network so that the other users can easily discover such items via social browsing, and (2) how to place items on a small fraction of users so that as many users as possible can search for such items by social browsing. Let us consider a more concrete application in Facebook social network. Assume that an application developer wants to popularize his Facebook application. Then, he may select a small fraction of users, say $k$ users, to install his application for free. Note that in Facebook, if a user has installed an application, then his friends can view such an application by browsing his home-page (social browsing). Therefore, the question is that how to select $k$ users so that the other users can easily find such an application (or as many users as possible can find such an application) which is equivalent to the question (1) (question (2)). Since we model the social browsing process as an $L$-length random walk, these questions are actually two instances of the random-walk domination problems.

\stitle{Optimizing Ads-placement in advertisement networks}: Similar example is also encountered in online advertisement networks, where an advertisement developer would like to place an advertisement (Ad) on a small fraction of users (he may pay for these users) such that it can be easily found by other users via social browsing (or as many users as possible can find such an Ad by social browsing). Likewise, we can model the user information-finding process in the advertisement networks as an $L$-length random walk. As a consequence, these problems become two instances of the random-walk domination problems.

\stitle{Accelerating resource search in P2P networks}: The study of the random-walk domination problems could also be beneficial to accelerate resource search in P2P networks. Specifically, in P2P network, how to place resources on a small number of peers such that other peers can easily search for such resources via some pre-specified search strategies. In P2P networks, a commonly-used search strategy is based on random walk \cite{06p2psearch}. Moreover, a resource-search process in P2P networks typically has a lifespan. That is to say, the resource-search process generally has a time or hops limit. Therefore, we can also model the resource-search process in P2P network as an $L$-length random walk, i.e., the resource-search process searches at most $L$ peers in its lifespan. Clearly, based on the $L$-length random walk, the resource placement problem in P2P network is an instance of the random-walk domination problem. Therefore, using the results of the random-walk domination problems can accelerate the resource search in P2P networks.

\subsection{Our main contributions}
This paper present the first study on the random-walk domination problems. Our goal is to formulate the random-walk domination problems and devise efficient and effective algorithms for these problems which can be directly applied to all the above applications. In particular, we first formulate two types of random-walk domination problems described above as two discrete optimization problems respectively. Then, we prove that these two problems are the instance of submodular set function maximization with cardinality constraint problem \cite{78submodular}. In general, such problems are NP-hard \cite{78submodular}. Therefore, we resort to develop approximate algorithms to solve them efficiently. To this end, we devise a dynamical programming (DP) based greedy algorithm to solve these problems effectively. By a well-known result \cite{78submodular}, the DP-based greedy algorithm achieves a $1-1/e$ ($\approx 0.63$) approximation factor. However, the time complexity of the DP-based greedy algorithm is over cubic w.r.t.\ the network size, thus it can only work well in the small graphs. To overcome this drawback, we develop an approximate greedy algorithm based on a carefully designed random-walk sampling and sample materialization techniques. The time and space complexity of the approximate greedy algorithm are linear w.r.t.\ the graph size, thereby it can be scalable to handle large graphs. Moreover, we show that the approximate greedy algorithm is able to achieve a $1-1/e-\epsilon$ approximation factor, where $\epsilon$ is a very small constant. Finally, we conduct comprehensive experiments over both synthetic and real-world graph datasets. The results indicate that the approximate greedy algorithm achieves very similar performance as the DP-based greedy algorithm, and it substantially outperforms the other baselines. In addition, the results demonstrate that the approximate greedy algorithm scales linearly w.r.t.\ the graph size.

The rest of this paper is organized as follows. Below, we will briefly review the existing studies that are related to ours. After that, we formulate the random-walk domination problems in Section~\ref{sec:modelproblem}. We propose the DP-based greedy algorithm and the approximate greedy algorithm for solving the random-walk domination problems in Section~\ref{sec:algs}. Extensive experiments are reported in Section~\ref{sec:experiments}. We conclude this work in Section \ref{sec:concl}.

\subsection{Related work}
Our problems are closely related to the dominating set problem in graphs. Dominating set problem is a classic NP-hard problem which has been well-studied in the literature \cite{98dominate1, 98dominate2}. There is an $O(\log n)$ approximate algorithm for solving this problem efficiently \cite{98dominate2}. Moreover, it has turned out that such an approximation factor is optimal unless P=NP \cite{98dominate2, 98jacmsetcover}. The dominating set problem has been widely-studied in the networking community due to a large number of applications in wireless sensor networks \cite{08mobihocdomset,04tpdsdominateset,07ccjdominate} and other Ad Hoc networks \cite{06tpdsextenddomset,08adhocdominate}. Recently, many different extensions of the dominating set problem have also been investigated. Notable examples include the distributed dominating set problem \cite{03podcdisdominateset}, the connected dominating set problem \cite{79connecteddomset, 98connecteddominateset, 04tpdsdominateset,06tpdsextenddomset}, the Steiner connected dominating set problem \cite{98connecteddominateset}, and the $k$-dominating set problem \cite{98dominate2,08mobihocdomset}. All of these extensions are based on the traditional definition of domination \cite{98dominate1} where the nodes deterministically dominate their immediate (or $L$-hop) neighbors. In our work, the problems are based on a newly defined concept called random-walk domination in which the targeted nodes dominate an $L$-hop neighbor if and only if such a neighbor-node hits one of the targeted nodes through an $L$-length random walk.

Our work is also related to the submodular set function maximization problem \cite{78submodular}. In general, the problem of submodular function maximization subject to cardinality constraint is NP-hard. Nemhauser et al.\ \cite{78submodular} propose a greedy algorithm with $1-1/e$ approximation factor to settle this issue. Recently, many applications are formulated as the submodular set function maximization subject to cardinality constraint problem. Some notable examples include the classic maximal $k$ coverage problem \cite{98jacmsetcover}, the influence maximization problem in social networks \cite{03kddinfluence}, the outbreak detection problem in networks \cite{07kddoutbreak}, the observation selection and sensor placement problem \cite{07aaaiobserselection, 08jmlrsensorplacement}, the document summarization problem \cite{10htldocsummarize, 11aclsubmodular}, the privacy preserving data publishing problem \cite{08aaiprivacy}, the diversified ranking problem
\cite{11icdmdivranklrh, 12tkdedivranklrh}, and the filter-placement problem \cite{12pvldbfilterplacement}. All of these problems can be approximately solved by the greedy algorithm given in \cite{78submodular}. Here we study two random-walk domination problems in graphs, and we show that both of them can also be formulated as the submodular set function maximization with cardinality constraint problem. Also, we present a near-optimal approximate greedy algorithm to solve them efficiently.

\section{Problems Statement}
\label{sec:modelproblem}
Consider an undirected and un-weighted graph $G=(V, E)$, where $V$ denotes a set of nodes and $E$ denotes a set of undirected edges. Let $n=|V|$ and $m=|E|$ be the number of nodes and the number of edges in $G$ respectively.  Although we only focus on undirected and un-weighted graphs in this paper, the proposed techniques can also be easily extended to directed and weighted graphs. Below, we first introduce some important concepts about random walk on graphs, and then we formulate two different types of random-walk domination problems.

\begin{figure}[t]
\begin{center}
\includegraphics[width=0.5\hsize]{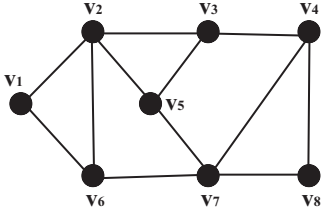}
\end{center}\vspace*{-2em}
\caption[]{Running example.}
\label{fig:expgraph} \vspace*{-0.5cm}
\end{figure}

A random walk on an undirected and un-weighted graph denotes the following process. Given an undirected and un-weighted graph $G$ and a starting node $u$, the random walk picks a neighbor-node of $u$ uniformly at random and moves to this neighbor-node, and then follows this way recursively \cite{93randomwalkgraph}. In this work, we address to a general random walk model called $L$-length random walk, where the path-length of the random walk is bounded by $L$ \cite{07uaihittingt}. It is important to note that the traditional random walk is a special case of the  $L$-length random walk by setting the parameter $L$ to infinity. Moreover, as discussed in Section~\ref{sec:intro}, many practical applications should be modeled by the $L$-length random walk. Let us consider a graph shown in Fig.~\ref{fig:expgraph}. Assume that $L=4$. Then, two possible paths generated by an $L$-length random walk starting from $v_1$ are $(v_1, v_2, v_3, v_2, v_6)$ and $(v_1, v_6, v_2, v_3, v_5)$. In which, both of them have a length $4$. Notice that the nodes could be repeatedly visited by the $L$-length random walk. For instance, in the first path, $v_2$ is visited twice by the $L$-length random walk.

Next, we define an important concept called hitting time for the $L$-length random walk. In particular, the hitting time between a source and targeted node measures the expected number of hops taken by an $L$-length random walk which starts at the source node and ends at the targeted node for the first time. Formally, denote by $Z_u ^t$ the position of an $L$-length random walk starting from node $u$ at discrete time $t$. Let $T ^L_{uv}$ be a random variable defined as
\begin{equation} \small \label{eq:hittimedef}
T^L_{uv} = \min \{\min\{t: Z_u ^t = v, t \ge 0\}, L\}.
\end{equation}
Then, the hitting time between node $u$ and $v$ denoted by $h_{uv} ^L$ is defined by the expectation of $T^L_{uv}$, i.e., $h_{uv} ^L = \mathbb{E}[T^L_{uv}]$. By this definition, the following lemma immediately holds.

\begin{lemma}
  \label{lem:hittimebound}
  For any two nodes $u$ and $v$, the hitting time $h_{uv} ^L$ is bounded by $L$, i.e., $h_{uv} ^L = \mathbb{E}[T^L_{uv}] \le L$.
\end{lemma}

The following theorem shows that the exact hitting time between two nodes can be computed recursively.

\begin{theorem}Let $d_u$ be the degree of node $u$ and $p_{uw}=1/d_u$ be the transition probability. Then,
for any nodes $u$ and $v$, $h_{uv}^L$ can be recursively computed by
  \label{thm:trunhitting}
\begin{equation} \small \label{eq:trunhitting}
h_{uv}^L  = \left\{ \begin{array}{l} 0,u = v \\
 1 + \sum\limits_{w \in V} {p_{uw} h_{wv}^{L - 1} } ,u \ne v, \\
 \end{array} \right.
 \end{equation}
 where $h_{wv}^{L - 1}$ denotes the hitting time between $w$ and $v$ based on an $(L-1)$-length random walk.
\end{theorem}

\begin{myproof}
  See Appendix.
\end{myproof}
\eop

 We remark that in \cite{07uaihittingt}, Sarkar and Moore define the hitting time of the $L$-length random walk in a recursive manner which is given in Eq.~(\ref{eq:trunhitting}). Note that our definition is more intuitive than their definition because our definition is based on Eq.~(\ref{eq:hittimedef}) in which the hitting time is ``explicitly'' bounded by $L$. In the above theorem, we show that our definition of hitting time can be computed by the same recursive equation (Eq.~(\ref{eq:trunhitting})) as defined in \cite{07uaihittingt}. Furthermore, based on Eq.~(\ref{eq:hittimedef}), it is very easy to design a sampling-based algorithm to estimate the hitting time. We will illustrate this point in Section~\ref{sec:algs}.

\subsection{The random-walk domination problems}
Based on the $L$-length random walk model, we introduce two types of random-walk domination problems in graphs. First, we describe the first type of random-walk domination problem as follows. Denote by $S \subseteq V$ a subset of nodes. Assume that there is an $L$-length random walk starting from a node $u \in V$. If such a random walk reaches any node in $S$ at any discrete time in $[0, L]$, we call that $u$ hits $S$ or $S$ dominates $u$ by an $L$-length random walk. For example, consider a graph shown in Fig.~\ref{fig:expgraph}. Suppose that $S=\{v_5, v_6\}$ and $L=4$. There is an $L$-length random walk $(v_1, v_2, v_3, v_2, v_6)$ starting from $v_1$. Since this random walk reaches node $v_6$ and $v_6 \in S$, we call that $v_1$ hits $S$ or $S$ dominates $v_1$. Clearly, if $u \in S$, then $u$ hits $S$. Next, we define another important concept called \emph{generalized hitting time} which measures the hitting time from a single source node to a set of targeted nodes $S$. Specifically, let $T^L_{uS}$ be a random variable defined as
\begin{equation} \small \label{eq:genhittimerv}
T^L_{uS} = \min \{\min\{t: Z_u ^t \in S, t \ge 0\}, L\}.
\end{equation}
By this definition, $T^L_{uS}$ denotes the number of hops of that the $L$-length random walk starting at $u$ hits any node in $S$ for the first time.
Reconsider the example in Fig.~\ref{fig:expgraph}. Suppose that $u=v_1$, $S=\{v_5, v_6\}$ and a $4$-length random walk $(v_1, v_2, v_3, v_2, v_6)$ starting at $v_1$. Then, $T^L_{uS}=4$ because the $L$-length random walk starting at $u=v_1$ hits a node $v_6 \in S$ at time 4 for the first time. Note that if $S=\emptyset$, we have $T^L_{uS}=L$. This is because if $S$ is an empty set, then $u$ cannot hit $S$, and thereby $\min\{t: Z_u ^t \in S, t \ge 0\}$ is infinity. In addition, if $L=0$, then $T^L_{uS}=0$ as $\min\{t: Z_u ^t \in S, t \ge 0\} \ge 0$. Based on $T^L_{uS}$, the generalized hitting time from $u$ to $S$ denoted by $h_{uS} ^L$ is defined by the expectation of $T^L_{uS}$, i.e., $h_{uS} ^L = \mathbb{E}[T^L_{uS}]$. By this definition, the smaller $h_{uS} ^L$ suggests that the node $u$ is more easier to hit a node in $S$ through an $L$-length random walk. Similarly, the generalized hitting time can be computed according to the following theorem.

\begin{theorem}
  \label{thm:ghitrur}
  For any node $u$ and set $S$, $h_{uS}^L$ can be computed by
  \begin{equation} \small\label{eq:ghitrur}
h_{uS}^L  = \left\{ \begin{array}{l} 0,u \in S \\
 1 + \sum\nolimits_{w \in V \backslash S} {p_{uw} h_{wS}^{L - 1} } ,u \notin S. \\
 \end{array} \right.
 \end{equation}
\end{theorem}

\begin{myproof}
  The proof is very similar to the proof of Theorem~\ref{thm:trunhitting}, thus we omit it for brevity.
\end{myproof}
\eop

Note that for $L=0$, we have $h_{uS} ^L=0$, as $T^0_{uS}=0$. Based on the generalized hitting time, the first type of random-walk domination problem is to minimize the sum of the generalized hitting time from the nodes in $V \backslash S$ to the targeted set of nodes $S$ subject to that $|S| \le k$.  More formally, this problem is formulated as
\begin{equation} \small\label{eq:problem2}
\begin{array}{l}
 \min \sum\limits_{u \in V\backslash S} {h_{uS}^L }  \\
 s.t. \quad |S| \le k. \\
 \end{array}
 \end{equation}

 It is easy to verify that the above optimization problem is equivalent to the following one. For convenience, in the rest of this paper, we refer to the following problem as the first type of random-walk domination problem and denoted it by Problem (1).

 \stitle{Problem (1):}
 \begin{equation} \small\label{eq:problem1r}
\begin{array}{l}
 \max nL - \sum\limits_{u \in V\backslash S} {h_{uS}^L }  \\
 s.t. \quad |S| \le k. \\
 \end{array}
 \end{equation}

Second, we formulate the second type of random-walk domination problem. Let $X_{uS}^L$  be a random variable such that $X_{uS}^L=1$ if node $u$ hits $S$ by an $L$-length random walk, $X_{uS}^L=0$ otherwise. Given a graph $G$ and a constant $k$, the second type of random-walk domination problem is to maximize the expected number of nodes that can be dominated by the set $S$ subject to a cardinality constraint, i.e., $|S| \le k$. Formally, the problem is defined as

\stitle{Problem (2):}
 \begin{equation} \small\label{eq:problem2}
 \begin{array}{l}
 \max \mathbb{E}[\sum\limits_{u \in V } {X_{uS}^L } ] \\
 s.t. \quad |S| \le k. \\
 \end{array}
 \end{equation}

Let $p_{uS}^L $ be the probability of an event that an $L$-length random walk starting from node $u$ successfully hits a node in $S$. Then, we have $\mathbb{E}[X_{uS} ^L] = p_{uS}^L $. Moreover, by definition, we have the following theorem.

\begin{theorem} For $L>0$, we have
  \label{thm:probtheorem}
  \begin{equation} \label{eq:probbiter2}\small
p_{uS}^L  = \left\{ \begin{array}{l}
 1,u \in S \\
 \sum\limits_w {p_{uw} p_{wS}^{L - 1} } ,u \notin S. \\
 \end{array} \right.
\end{equation}
\end{theorem}

\begin{myproof}
  The proof can be easily obtained by definition, we therefore omit for brevity.
\end{myproof}
\eop

For $L=0$, we define $p_{uS}^0=1$ if $u \in S$, $p_{uS}^0=0$ otherwise. The rationale is that a 0-length random walk means that a node does not walk to any other nodes. Therefore, if $u \in S$, we have $p_{uS}^0=1$, $p_{uS}^0=0$ otherwise. It is important to emphasize that Problem (2) is different from Problem (1). Because Problem (2) is to maximize the expected number of nodes that hit the targeted set by the $L$-length random walk, while Problem (1) is to minimize the total expected time (or the expected number of hops) of which every node hits the targeted set.

\stitle{Distinguishing Problem (2) from the influence maximization problems}: The influence maximization problem in social networks is to select $k$ nodes to maximize the expected influence spread from those $k$ nodes based on a influence spread model \cite{03kddinfluence}. A commonly-used influence spread model is the independent cascade model \cite{03kddinfluence}, where a user influences his friends with a pre-specified probability and the influence spread along an edge is independent of the influence spread over the other edges. More specifically, under the independent cascade model, the social network is modeled by a probabilistic graph, where each edge is associated with a probability and all of those probabilities are independent of one another. The influence maximization problem is to select $k$ nodes to maximize the expected number of nodes that are reachable from the selected nodes. Recall that Problem (2) is to select $k$ nodes to maximize the expected number of nodes that can reach a node in the targeted node set following an $L$-length random walk. Although these two problems are seemingly similar, the Problem (2) is totally different from the influence maximization problem. The reasons are as follows. First, Problem (2) is based on an $L$-length random walk model which is a Markov-Chain model where the visiting probability of a node depends on the visiting probability of its immediate neighbors. The influence maximization problem, however, is based on the independent cascade model where the probabilities associated on the edges are independent of one another. Second, in the influence maximization problem, a targeted node could influence multiple immediate neighbors at a discrete time. However, in an $L$-length random walk model, each node only follows one immediate neighbor. Let us consider a concrete example to illustrate this point. For example, in Fig.~\ref{fig:expgraph}, we assume that there is a $4$-length random walk $(v_1, v_2, v_3, v_2, v_6)$ starting from $v_1$. Suppose that in the independent cascade model, the node $v_1$ has successfully influenced node $v_2$ and $v_3$. Clearly, in this case, $v_1$ has only one descendant node in the $L$-length random walk model, while in the independent cascade model $v_1$ has two. Finally, the influence maximization problem relies on the predefined influence probabilities where all the influence probabilities are the input parameters. In Problem (2), we do not require the knowledge of influence probabilities. The only input parameters of our problems are the graph topology and the parameter $k$.


\section{The algorithms}
\label{sec:algs} The goal of this section is to present algorithmic treatments for Problem (1) and Problem (2). Specifically, we first prove that both Problem (1) and Problem (2) are the instances of the submodular set function maximization with cardinality constraint problem \cite{78submodular}. In general, these problems are NP-hard \cite{78submodular}. Therefore, we strive to devise approximate algorithms for these problems. In the following, we will present two efficient greedy algorithms for Problem (1) and Problem (2) with near-optimal performance guarantee.

\subsection{Submodularity and greedy algorithm}\label{subsec:greedyalg}
Before we proceed, let us give a definition of the non-increasing submodular set function \cite{78submodular}.

\begin{definition}
\label{def:submodular} Let $V$ be a finite set, a real valued function $f(S)$ defined on the subsets of $V$, i.e, $S \subseteq V$, is called a
nondecreasing submodular set function, if the following conditions hold.
\begin{itemize}
\item \textbf{Nondecreasing}: For any subsets $S$ and $T$ of
$V$ such that $S \subseteq T \subseteq V$, we have $f(S) \le f(T)$.

\item \textbf{Submodularity}: Let $\sigma _j (S) = f(S \cup \{ j\} ) - f(S)$ be the marginal gain. Then, for any subsets $S$ and $T$ of $V$ such that $S \subseteq T \subseteq V$ and $j \in V \backslash T$, we have $\sigma _j (S) \ge \sigma _j (T)$.
\end{itemize}
\end{definition}

Then, based on the definition of submodular function, we show that the objective functions of Problem (1) and Problem (2) are submodular. Specifically, let {\small $F_1(S) = nL - \sum\nolimits_{u \in V\backslash S} {h_{uS}^L}$}, and {\small $F_2(S)=\mathbb{E}[\sum\nolimits_{u \in V } {X_{uS}^L } ]$}.  Then, we have the following two theorems.

\begin{theorem}\label{thm:f1submodular}
$F_1(S)$ is a non-increasing submodular set function with $F_1(\emptyset) = 0$.
\end{theorem}

\begin{myproof}
  See Appendix.
\end{myproof}
\eop

\begin{theorem}\label{thm:f2submodular}
$F_2(S)$ is a non-increasing submodular set function with $F_2(\emptyset)
= 0$.
\end{theorem}

\begin{myproof}
  See Appendix.
\end{myproof}
\eop

\begin{algorithm}[t]
\caption{The greedy algorithm}
\label{alg:exact}
 {\small
\begin{tabbing}
    {\bf\ Input}: \hspace{0.3cm}\= A graph $G=(V, E)$, and a parameter $k$\\
{\bf\ Output}: \> A set of nodes $S$
\end{tabbing}
\begin{algorithmic}[1]

\STATE $S \leftarrow \emptyset$;
\FOR {$i = 1$ to $k$}
    \STATE $v \leftarrow \arg \mathop {\max }\limits_{u \in V \backslash S} \{ F(S \cup \{ u\} ) - F (S)\} $;
    \STATE $S \leftarrow S \cup \{ v\} $;
\ENDFOR
\STATE \textbf{return} $S$;
\end{algorithmic}
}
\end{algorithm}

Based on the submodularity of $F_1$ and $F_2$, we present a greedy algorithm for both Problem (1) and Problem (2) depicted in Algorithm~\ref{alg:exact}. The greedy algorithm works in $k$ rounds (line~2-4). In each round, the algorithm selects a node with maximal marginal gain (line~3), and adds it into the answer set $S$ (line~4) which is initialized by an empty set (line~1). Note that to solve the Problem (1) and Problem (2), we need to replace the function $F$ in Algorithm~\ref{alg:exact} with $F_1$ and $F_2$ respectively. By a celebrated result in \cite{78submodular}, Algorithm~\ref{alg:exact} achieves a $(1-1/e)$ approximation factor for problem (1) and problem (2), where $e \approx 2.718$ denotes the Euler's number.

\stitle{Complexity analysis}: The time complexity of Algorithm~\ref{alg:exact} is dominated by the time complexity for computing the marginal gain (line~3). Below, we focus on an analysis of the greedy algorithm for Problem (1), and similar analysis can be used for Problem (2). For $F_1$, let $\sigma_u(S)  = F_1 (S) - F_1 (S \cup \{ u\} )$ be the marginal gain. Then, $\sigma_u(S)$ can be calculated based on Eq.~(\ref{eq:ghitrur}). Note that Eq.~(\ref{eq:ghitrur}) immediately implies a dynamic programming algorithm for computing $h_{uS}^L$. Given a set $S$, the time complexity for computing $h_{uS}^L$ is $O(mL)$. Therefore, given a set $S$, the time complexity for calculating $F_1(S) = \sum\nolimits_{u \in V\backslash S} {(L-h_{uS}^L)}$ is $O(nmL)$. Since the greedy algorithm needs to find the node with maximal marginal gain, it needs to evaluate $F_1(S \cup \{ u\} )$ for every node $u$ in $V \backslash S$. As a result, the time complexity of the greedy algorithm is $O(kn^2mL)$. We can use the so-called lazy evaluation strategy \cite{07kddoutbreak} to speed up the greedy algorithm, which could result in several orders of magnitude speedup as observed in \cite{07kddoutbreak}. For the space complexity, the dynamic programming algorithm needs to maintain a $n \times L$ array for a given $S$. To compute the marginal gain, the greedy algorithm needs to evaluate $F_1(S \cup \{ u\} )$ for every node $u$ in $V \backslash S$, thus the space complexity of the greedy algorithm is $O(n^2L)$. Similarly, for problem (2), the time and space complexity of the greedy algorithm are $O(kn^2mL)$ and $O(n^2L)$ respectively.

\stitle{Approximate marginal gain computation}: Based on the complexity analysis, the greedy algorithm is clearly impractical. The most time and space consuming step in the greedy algorithm is to compute the objective functions and the corresponding marginal gains. Here we present a sampling-based algorithm to approximately compute the objective functions and the marginal gains efficiently.

Given a set $S$, to estimate the objective function $F_1(S)$ ($F_2(S)$), the key step is to estimate $h_{uS} ^L$ (${\mathbb{E}}[X_{uS}^L]$). Below, we firstly describe an unbiased estimator for estimating $h_{uS} ^L$. To construct an unbiased estimator for $h_{uS}^L$, we independently run $R$ $L$-length random walks starting from node $u$. Assume that there are $r$ such random walks that have hit any arbitrary node in $S$ for the first time at $\{t_{i_1}, \cdots, t_{i_r}\}$ hops. Then, we construct an estimator for $h_{uS}^L$ by
\begin{equation} \small
\hat h_{uS}^L  = \frac{{\sum\nolimits_{k = 1}^r {t_{i_k } } }}{R} +
(1 - \frac{r}{R})L.
\end{equation}

The following lemma shows that $\hat h_{uS}^L$  is an unbiased estimator of $h_{uS}^L$.

\begin{lemma}
  \label{lem:unbiasede1}
  $\hat h_{uS}^L$ is an unbiased estimator of $h_{uS}^L$.
\end{lemma}

\begin{myproof}
Recall that $h_{uS} ^L = \mathbb{E}[T^L_{uS}]$. By Eq.~(\ref{eq:genhittimerv}), $T^L_{uS}$ denotes the first time that an $L$-length random walk starting from $u$  hits any arbitrary node in $S$. If such a random walk cannot hit the nodes in $S$, then $T^L_{uS}=L$. To estimate the expectation of $T^L_{uS}$, we independently run $R$ $L$-length random walks starting from $u$, and take the average hitting time as the estimator. The proposed sampling process is equivalent to a simple random sampling with replacement, thus the estimator is unbiased.
\end{myproof}
\eop

Based on $\hat h_{uS}^L$ and Lemma~\ref{lem:unbiasede1}, $\hat F_1(S) = \sum\nolimits_{u \in V\backslash S} {(L-\hat h_{uS}^L)}$ is also an unbiased estimator of $F_1(S)$. Similarly, we can construct an estimator for ${\mathbb{E}}[X_{uS}^L]$ by
\begin{equation} \small
\hat {\mathbb{E}}[X_{uS}^L] = \frac{r}{R}.
\end{equation}

Also, the estimator $\hat {\mathbb{E}}[X_{uS}^L]$ is unbiased.

\begin{lemma}
  \label{lem:unbiasede2}
  $\hat {\mathbb{E}}[X_{uS}^L]$ is an unbiased estimator of ${\mathbb{E}}[X_{uS}^L]$.
\end{lemma}

\begin{myproof}
  The proof can be easily obtained by definition, we omit it for brevity.
\end{myproof}
\eop

Likewise, based on {\small $\hat {\mathbb{E}}[X_{uS}^L]$} and Lemma~\ref{lem:unbiasede2}, {\small $\hat F_2(S)=\sum\nolimits_{u \in V } \hat {\mathbb{E}}[X_{uS}^L]$} is an unbiased estimator of {\small $F_2(S)$}. We remark that in \cite{08icmlhittingt}, Sarkar et al.\ presented a similar unbiased estimator for estimating the hitting time of the $L$-length random walk between two nodes. Here our estimator ($\hat h_{uS}^L$) is to estimate the hitting time of the $L$-length random walk between one source node and one targeted set. In this sense, our estimator is more general than the estimator presented in \cite{08icmlhittingt}. Below, we make use of the Hoeffding inequality \cite{63hoeffdingbound} to bound the sample size $R$. Specifically, we have the following two lemmas.


\begin{lemma} \label{lem:estimator1bound}
Given a set $S$, for two small constants $\epsilon$ and $\delta$, if $R \ge \frac{1}{{2\varepsilon ^2 }}\log \frac{n-|S|}{\delta }$, then $\Pr[|\hat F_1(S)-F_1(S)| \ge \epsilon (n-|S|)L] \le \delta$.
\end{lemma}

\begin{myproof}
First, we have 
\[ \small
\begin{array}{l}
 \Pr [|\hat F_1 (S) - F_1 (S)| \ge \varepsilon (n - |S|)L] \\
  \quad \quad \quad \quad \le \Pr [\sum\nolimits_{u \in V/S} {|\hat h_{uS}  - h_{uS} |}  \ge \varepsilon (n - |S|)L], \\
 \end{array}
\]
\noindent
because the event of $|\hat F_1 (S) - F_1 (S)| \ge \varepsilon (n-|S|)L$ implies the event of $\sum\nolimits_{u \in V/S} {|\hat h_{uS}  - h_{uS} | \ge \varepsilon (n-|S|)L}$. Then, by the union bound, we have
\[ \small
\begin{array}{l}
 \Pr [\sum\limits_{u \in V/S} {|\hat h_{uS}  - h_{uS} |}  \ge \varepsilon (n - |S|)L] \\
  \quad \quad \quad \quad \le \sum\nolimits_{u \in V/S} {\Pr [|\hat h_{uS}  - h_{uS} | \ge \varepsilon L. ]}  \\
 \end{array}
\]

Since $0 \le \hat h_{uS} \le L$ (Lemma~\ref{lem:hittimebound}), we can apply the Hoedding inequality \cite{63hoeffdingbound} to bound sample size $R$. Specifically, we have
\[ \small
\Pr [|\hat h_{uS}  - h_{uS} | \ge \varepsilon L] \le \exp ( - 2\varepsilon ^2 R).
\]
Based on this, the following inequality immediately holds
\[ \small
\Pr [|\hat F_1 (S) - F_1 (S)| \ge \varepsilon (n - |S|)L] \le (n - |S|)\exp ( - 2\varepsilon ^2 R).
\]
Let $(n - |S|)\exp ( - 2\varepsilon ^2 R) \le \delta$, then we can get $R \ge \frac{1}{{2\varepsilon ^2 }}\log \frac{n-|S|}{\delta }$, which completes the proof.
\end{myproof}
\eop

\begin{lemma}\label{lem:estimator2bound}
  Given a set $S$, for two small constants $\epsilon$ and $\delta$, if $R \ge \frac{1}{{2\varepsilon ^2 }}\log \frac{n}{\delta }$, then $\Pr[|\hat F_2(S)-F_2(S)| \ge \epsilon n] \le \delta$.
\end{lemma}

\begin{myproof}
  The proof is similar to the proof of Lemma~\ref{lem:estimator1bound}, thus we omit for brevity.
\end{myproof}
\eop

Based on the above analysis, in Algorithm~\ref{alg:sampling}, we present a sampling-based algorithm to estimate $F_1(S)$ and $F_2(S)$ given a set $S$. Note that the marginal gains $\sigma_u(S)=F_1(S \cup \{u\}) - F_1(S)$ and $\rho_u(S)=F_2(S \cup \{u\}) - F_2(S)$ can be easily estimated by invoking Algorithm~\ref{alg:sampling} twice. There are three input parameters $L$, $R$, and $S$ in Algorithm~\ref{alg:sampling}, where $R$ is a small value and it can be determined according to Lemma~\ref{lem:estimator1bound} and Lemma~\ref{lem:estimator2bound}. To compute the estimator of $\hat F_1(S)$ and $\hat F_2(S)$, for each node in $V \backslash S$, Algorithm~\ref{alg:sampling} independently runs $R$ $L$-length random walks (line~3-15), and records two quantities $r$ and $t$ (line~9-11). Based on $r$ and $t$,  Algorithm~\ref{alg:sampling} can easily compute $\hat F_1(S)$ and $\hat F_2(S)$ (line~12-15). It is worth mentioning that for the node $u \in S$, we have ${\mathbb{E}}[X_{uS}^L]=1$. Therefore, in line~15, the algorithm adds $|S|$ into $\hat F_2(S)$. Finally, the algorithm outputs the two estimators.

\begin{algorithm}[t]
\caption{Sampling algorithm for estimating $F_1(S)$ and $F_2(S)$}
\label{alg:sampling}
 {\small
\begin{tabbing}
    {\bf\ Input}: \hspace{0.3cm}\= A graph $G=(V, E)$, two parameters $L$ and $R$\\
    \> and a set $S$ \\
{\bf\ Output}: \>  Unbiased estimators for $\hat F_1(S)$ and $\hat F_2(S)$
\end{tabbing}
\begin{algorithmic}[1]
\STATE $\hat F_1(S) \leftarrow 0$;
\STATE $\hat F_2(S) \leftarrow 0$;
\FOR {each node $u \in V \backslash S$}
    \STATE $r \leftarrow 0$;
    \STATE $t \leftarrow 0$;
    \FOR {$i=1:R$}
        \STATE Run an $L$-length random walk from $u$;
        \IF{the random walk hits any arbitrary node $v$ in $S$ for the first time}
            \STATE $r \leftarrow r + 1$;
            \STATE Record $t_i$ be the number of nodes of the random walk segment from node $u$ to $v$;
            \STATE $t \leftarrow t + t_i$;
        \ENDIF
    \ENDFOR
    \STATE $\hat F_1(S) \leftarrow F_1(S) + (t+(R-r)L)/R$;
    \STATE $\hat F_2(S) \leftarrow F_2(S) + r / R$;
\ENDFOR
\STATE $\hat F_1(S) \leftarrow |V \backslash S| \times L - \hat F_1(S)$;
\STATE $\hat F_2(S) \leftarrow \hat F_2(S) + |S|$;
\STATE \textbf{return} $\hat F_1(S)$ and $\hat F_2(S)$;
\end{algorithmic}
}
\end{algorithm}

The time complexity of Algorithm~\ref{alg:sampling} is $O(nRL)$. This is because, running an $L$-length random walk takes $O(L)$ time complexity, and for each node, the algorithm needs to run $R$ $L$-length random walks. The space complexity of Algorithm~\ref{alg:sampling} is $O(m+n)$, which is linear w.r.t.\ the graph size. Based on Algorithm~\ref{alg:sampling}, the time complexity of the greedy algorithm is reduced to $O(kn^2RL)$, and the space complexity of the greedy algorithm is linear, which is significantly better than the greedy algorithm with exact marginal gain computation using a dynamic programming (DP) algorithm. Since Algorithm~\ref{alg:sampling} can be applied to compute a good approximation of the marginal gain, the performance guarantee of the greedy algorithm with sampling-based marginal gain computation can be preserved. In effect, by a similar analysis presented in \cite{03kddinfluence}, such a greedy algorithm can achieve a $1-1/e-\epsilon$ approximation factor through setting an appropriate parameter $R$. 
In addition, it is worth noting that the sampling-based greedy algorithm can also be accelerated using the lazy evaluation strategy \cite{07kddoutbreak}.

\subsection{Approximate greedy algorithm} \label{subsec:approxgreedy}
Although the sampling-based greedy algorithm are much more efficient than the DP-based greedy algorithm, the time complexity of the sampling-based greedy algorithm is $O(kn^2RL)$, which implies that such an algorithm can only be scalable to medium size graphs. Here we propose an approximate greedy algorithm for both problem (1) and problem (2) with linear time complexity (w.r.t.\ graph size) and near-optimal performance guarantee. Recall that in the sampling-based greedy algorithm, we need to invoke the sampling algorithm (Algorithm~\ref{alg:sampling}) to estimate the marginal gain $\sigma_u(S)$ for each node $u$. In each round, the greedy algorithm needs to find the node with maximal marginal gain. Note that there are $n-|S|$ nodes in total. Thus, the sampling-based greedy algorithm requires to invoke Algorithm~\ref{alg:sampling} $O(kn)$ times in $k$ rounds, which indicates that the algorithm needs to run $O(kn^2R)$ $L$-length random walks. Can we reduce the \emph{sample complexity} of the sampling-based greedy algorithm? In this subsection, we give an algorithm that only requires to run $O(nR)$ $L$-length random walks, and it also preserves the $1-1/e-\epsilon$ approximation factor. For convenience, we call this algorithm an approximate greedy algorithm. Below, we mainly focus on describing the algorithm for problem (1), and similar descriptions can be used for problem (2) (we have added some remarks for problem (2) in Algorithm~\ref{alg:indexcons}, \ref{alg:approxgain}, \ref{alg:update}, and \ref{alg:approxgreedy}).

The key idea is described as follows. First, for each node, the algorithm independently runs $R$ $L$-length random walks. Then, the algorithm materializes such \emph{samples} (An $L$-length random walk is a sample), and applies them to estimate the marginal gain $\sigma_u(S)$ for any given node $u$ and a given set $S$. Here the challenge is how to estimate $\sigma_u(S)$ efficiently using such samples, because $S$ changes in each round of the greedy algorithm. To overcome this challenge, we present an inverted list structure to index the samples. Specifically, we build $R$ inverted lists, and each inverted list includes $n$ sublists. For each node $u$, a sublist indexes all the other nodes that hit $u$ through an $L$-length random walk. Here the entry of the sublist is an object that includes two parts: a node ID ($id$) and a weight ($weight$), denoting $id$ hits $u$ at $weight$-th hop. Algorithm~\ref{alg:indexcons} depicts the inverted index construction algorithm. In Algorithm~\ref{alg:indexcons}, the $R$ inverted lists, denoted by $I[1:R][1:n]$, are organized as a two-dimensional list array, in which $I[i][v]$ indexes all the nodes that hit $v$ by the $i$-th $L$-length random walk. First, the algorithm initializes $I[1:R][1:n]$ by an empty array (line~1). Then, for each node $w$ in $V$, the algorithm runs $R$ $L$-length random walks (line~2-14). Let us consider the $i$-th $L$-length random walk starting at node $w$. If $w$ hits a node $v$, the algorithm creates an object $<w, weight>$, where $weight$ denotes that $w$ hits $v$ at $weight$-hop (line~11-12). Then, the algorithm adds it into $I[i][v]$ (line~13). Note that for the repeated nodes in an $L$-length random walk, we only need to index one node and record the $weight$ at the first visiting time according to the definition of hitting time. To remove such repeated nodes in an $L$-length random walk, the algorithm maintains a $visited[1:n]$ array (line~4, 6 and 9-10).

\begin{algorithm}[t]\caption{Invert\_Index($G$, $L$, $R$)}
\label{alg:indexcons}
 {\small
\begin{tabbing}
    {\bf\ Input}: \hspace{0.3cm}\= A graph $G=(V, E)$, two parameters $L$ and $R$\\
{\bf\ Output}: \>  An inverted index $I[1:R][1:n]$
\end{tabbing}
\begin{algorithmic}[1]
\STATE Initialize an inverted list $I[1:R][1:n] \leftarrow NULL$ ;
\FOR {each node $w \in V$}
    \FOR {$i=1:R$}
    \STATE Initialize $visited[1:n] \leftarrow 0$;
    \STATE $u \leftarrow w$;
    \STATE $visited[u] \leftarrow 1$;
        \FOR {$j=1:L$}
            \STATE Randomly select a neighbor of $u$, denoted by $v$;
            \IF {$visited[v] == 0$}
                \STATE $visited[v] \leftarrow 1$;
                \STATE $Object.id \leftarrow w$;
                \STATE $Object.weight\leftarrow j$; /*$w$ hits $v$ at $j$-th step*/\\
                 /*$Object.weight \leftarrow 1$; for problem (2)*/;
                \STATE $I[i][v].push\_back(Object)$;
            \ENDIF
            \STATE $u \leftarrow v$;
        \ENDFOR
    \ENDFOR
\ENDFOR
\STATE \textbf{return} $I[1:R][1:n]$;
\end{algorithmic}
}
\end{algorithm}

Given the inverted lists $I[1:R][1:n]$, how to estimate the marginal gain for any node $u$ and a given set $S$? Here we tackle this issue by maintaining a two-dimensional array $D[1:R][1:n]$. Given a set $S$, $D[i][u]$ denotes an estimator of the hitting time $h_{uS}^L$ based on the $i$-th $L$-length random walk.  Let $S_u=S \cup \{u\}$, and $\sigma _u (S)=F_1(S_u)-F_1(S)$ be the marginal gain. Then, we can derive that $\sigma _u (S) = \sum\nolimits_{w \in V\backslash S_u} {(h_{wS}^L -h_{wS_u }^L )}  + h_{uS}^L  - L$. Recall that in each round of the greedy algorithm, we need to find the node with maximal marginal gain. Therefore, for each node $u$, we can estimate $\sigma _u$ by $\sum\nolimits_{w \in V\backslash S_u} {(h_{wS}^L -h_{wS_u }^L)}  + h_{uS}^L$, because ``$-L$'' dose not affect the results. Algorithm~\ref{alg:approxgain} describes an algorithm for estimating $\sigma _u$. Let us consider the $i$-th $L$-length random walk. First, $\sigma _u$ is initialized by 0. Then, the algorithm adds $D[i][u]$, which is an estimator of $h_{uS}^L$, to $\sigma _u$ (line~3). And then, the algorithm estimates $\sum\nolimits_{w \in V\backslash S_u} {(h_{wS}^L -h_{wS_u }^L) }$ and adds it to $\sigma _u$, which is implemented in line~4-7. By definition, if a node $v$ in $V \backslash S_u$ dose not hit $u$, then we have $h_{vS}^L =h_{vS_u }^L$. Thus, the algorithm only needs to consider the nodes that hit $u$ (line~4), which is indexed in $I[i][u]$. If $h_{vu}^L < h_{vS}^L$, then the algorithm adds $h_{vS}^L-h_{vu}^L $ to $\sigma _u$. Otherwise, we have $h_{vS}^L =h_{vS_u }^L$. Note that by definition, $h_{vu}^L$ can be estimated by the $weight$ associated with $v$ which is indexed in $I[i][u]$, and $h_{vS}^L$ can be estimated by $D[i][v]$, and thus $h_{vS}^L-h_{vu}^L $ can be estimated by $D[i][v]$ minus the $weight$ associated with $v$ (line~7). Therefore, line~3-7 of Algorithm~\ref{alg:approxgain} is to estimate $\sigma _u$ based on the $i$-th $L$-length random walk. Finally, Algorithm~\ref{alg:approxgain} takes an average over all the $R$ estimators (line~10).

\begin{algorithm}[t]
\caption{Approx\_Gain($I[1:R][1:n]$, $D[1:R][1:n]$, $u$, $R$)}
\label{alg:approxgain}
 {\small
\begin{tabbing}
    {\bf\ Input}: \hspace{0.3cm}\= The inverted index $I[1:R][1:n]$, the array $D[1:R][1:n]$, \\
    \> a node $u$ and parameter $R$\\
{\bf\ Output}: \>  Approximate marginal gain $\sigma_u$
\end{tabbing}
\begin{algorithmic}[1]
\STATE Initialize $\sigma_u \leftarrow 0$;
\FOR {$i=1:R$}
    \STATE $\sigma_u  \leftarrow \sigma_u + D[i][u]$;\\
    /*$\sigma_u  \leftarrow \sigma_u + 1 - D[i][u]$; for problem (2)*/
    \WHILE{$Object \leftarrow I[i][u].pop()$}
        \STATE $v \leftarrow Object.id$;
        \IF {$Object.weight < D[i][v]$}
            \STATE $\sigma_u  \leftarrow \sigma_u + D[i][v] - Object.weight$;
        \ENDIF
        /*for problem (2), use line~8-9 to replace line~6-7*/
        \IF {$Object.weight > D[i][v]$}
            \STATE $\sigma_u  \leftarrow \sigma_u +  Object.weight - D[i][v]$;
        \ENDIF
    \ENDWHILE
\ENDFOR
 \STATE $\sigma_u \leftarrow \sigma_u / R$;
\STATE \textbf{return} $\sigma_u$;
\end{algorithmic}
}
\end{algorithm}

\begin{algorithm}[t]
\caption{Update($I[1:R][1:n]$, $D[1:R][1:n]$, $u$, $R$)}
\label{alg:update}
 {\small
\begin{tabbing}
    {\bf\ Input}: \hspace{0.3cm}\= The inverted index $I[1:R][1:n]$, the array $D[1:R][1:n]$, \\
    \> a node $u$ and parameter $R$\\
{\bf\ Output}: \>  The updated array $D[1:R][1:n]$
\end{tabbing}
\begin{algorithmic}[1]
\FOR {$i=1:R$}
    \STATE $D[i][u] \leftarrow 0$;
    /*$D[i][u] \leftarrow 1$; for problem (2)*/
    \WHILE{$Object \leftarrow I[i][u].pop()$}
        \STATE $v \leftarrow Object.id$;
        \IF {$Object.weight < D[i][v]$}
            \STATE $D[i][v]  \leftarrow Object.weight$;
        \ENDIF
         /*for problem (2), use line~7-8 to replace line~5-6)*/
        \IF {$Object.weight > D[i][v]$}
            \STATE $D[i][v] \leftarrow Object.weight$;
        \ENDIF
    \ENDWHILE
\ENDFOR
\end{algorithmic}
}
\end{algorithm}

\begin{algorithm}[t]
\caption{The approximate greedy algorithm}
\label{alg:approxgreedy}
 {\small
\begin{tabbing}
    {\bf\ Input}: \hspace{0.3cm}\= A graph $G=(V, E)$, and a parameter $k$\\
{\bf\ Output}: \> A set of nodes $S$
\end{tabbing}
\begin{algorithmic}[1]
\STATE $I[1:R][1:n]\leftarrow$Invert\_Index($G$, $L$, $R$);
\STATE $S \leftarrow \emptyset$;
\STATE Initialize $D[1:R][1:n] \leftarrow L$; \\ /*$D[1:R][1:n] \leftarrow 0$; for problem (2)*/
\FOR {$i = 1$ to $k$}
    \STATE $v \leftarrow \arg \mathop {\max }\limits_{u \in V \backslash S} \; $Approx\_Gain$(I[1:R][1:n], D[1:R][1:n], u, R) $;
    \STATE $S \leftarrow S \cup \{ v\} $;
    \STATE Update($I[1:R][1:n]$, $D[1:R][1:n]$, $v$, $R$);
\ENDFOR
\STATE \textbf{return} $S$;
\end{algorithmic}
}
\end{algorithm}

Algorithm~\ref{alg:approxgain} can be used to estimate the marginal gain for every node given a set $S$. In the greedy algorithm, after one round, the size of $S$ increases by 1. Hence, we need to dynamically maintain the array $D[1:R][1:n]$ when $S$ is changed. Algorithm~\ref{alg:update} depicts an algorithm to update $D[1:R][1:n]$ given $S$ is inserted an element $u$. As usual, let us consider the $i$-th $L$-length random walk. By definition, for a node $v$, if $h_{vu}^L < h_{vS}^L$, then we need to update $D[i][v]$. Otherwise, we have $h_{vS}^L=h_{vS_u}^L$, thus no need to update $D[i][v]$. In addition, for a node $v$ that does not hit $u$, we do not need to update $D[i][v]$ as $h_{vS}^L=h_{vS_u}^L$ by definition. In Algorithm~\ref{alg:update}, the algorithm firstly sets $D[i][u]$ to 0 (line~2), because $h_{uS_u}^L=0$ ($u$ is in $S_u$). Then, the algorithm updates $D[i][v]$ for the node $v$ that has hit $u$ by the $i$-th $L$-length random walk (line~3-6).

Equipped with Algorithm~\ref{alg:indexcons}, Algorithm~\ref{alg:approxgain}, and Algorithm~\ref{alg:update}, we present the approximate greedy algorithm in Algorithm~\ref{alg:approxgreedy}. First, Algorithm~\ref{alg:approxgreedy} builds $R$ inverted lists (line~1). Second, the algorithm initializes the answer set $S$ to an empty set (line~2), and sets the value of each entry in $D[1:R][1:n] $ to $L$ (line~3), because $h_{uS}^L=L$ given $S=\emptyset$. Third, the algorithm works in $k$ rounds (line~4-7). In each round, the algorithm invokes Algorithm~\ref{alg:approxgain} to estimate the marginal gain $\sigma_u(S)$, and selects the node $v$ with maximal $\sigma_u(S)$. Then, the algorithm adds $v$ into the answer set $S$. After that, the algorithm invokes Algorithm~\ref{alg:update} to update $D[1:R][1:n]$. The following example illustrates how the Algorithm~\ref{alg:approxgreedy} works.

\begin{example}
  Let us re-consider the example graph shown in Fig.~\ref{fig:expgraph}. For simplicity, we set $R=1$, $L=2$, and $k=2$. Suppose that the $2$-length random walks for each node are described as follows: $(v_1, v_2, v_3)$, $(v_2, v_3, v_5)$, $(v_3, v_2, v_5)$, $(v_4, v_7, v_5)$, $(v_5, v_2, v_6)$, $(v_6, v_7, v_5)$, $(v_7, v_5, v_7)$, and $(v_8, v_7, v_4)$. Then, the inverted index constructed by Algorithm~\ref{alg:indexcons} ($I[1][1:8]$) is illustrated Table~\ref{tbl:invertindex}.
  \begin{table}[htbp]
\begin{center}\vspace*{-0.5cm}\small
\caption[]{Inverted index} \label{tbl:invertindex}
\begin{tabular}{l|l}
\hline \hline
$v_1$: & \\
$v_2$: & $<v_1, 1>$, $<v_3, 1>$, $<v_5, 1>$ \\
$v_3$: & $<v_1, 2>$, $<v_2, 1>$ \\
$v_4$: & $<v_8, 2>$ \\
$v_5$: & $<v_2, 2>$, $<v_3, 2>$, $<v_4, 2>$, $<v_6, 2>$, $<v_7, 1>$ \\
$v_6$: & $<v_5, 2>$ \\
$v_7$: & $<v_4, 1>$, $<v_6, 1>$, $<v_8, 1>$ \\
$v_8$: & \\
\hline
\end{tabular}\vspace*{-0.5cm}
\end{center}
\end{table}
Note that in $(v_7, v_5, v_7)$, $v_7$ is a repeated node, thus the second $v_7$ will not be inserted into the inverted list by Algorithm~\ref{alg:indexcons}. After building the inverted index, Algorithm~\ref{alg:approxgreedy} initializes $S$ to an empty set, and set all the elements of $D[1][1:8]$ to 2. Then, in the first round, the algorithm invokes Algorithm~\ref{alg:approxgain} to estimate the marginal gain $\sigma_u(\emptyset)$ for each node. After this step, we can get that $\sigma_{v_1}(\emptyset)=2$, $\sigma_{v_2}(\emptyset)=5$, $\sigma_{v_3}(\emptyset)=3$, $\sigma_{v_4}(\emptyset)=2$, $\sigma_{v_5}(\emptyset)=3$, $\sigma_{v_6}(\emptyset)=2$, $\sigma_{v_7}(\emptyset)=5$, and $\sigma_{v_8}(\emptyset)=2$.
For instance, for node $v_2$, there are three elements in the inverted list $I[1][2]$. Since the weights of $v_1$, $v_3$, and $v_5$ (all of them equal to 1) are smaller than $D[1][1]$, $D[1][3]$, and $D[1][5]$ (all of them equal to 2) respectively, thus $\sigma_{v_2}(\emptyset)=D[1][2] + 3=5$ as desired. Similar analysis can be used for other nodes. Clearly, $v_2$ and $v_7$ achieve the maximal marginal gain. The algorithm breaks ties randomly. Assume that in this round, the algorithm selects $v_2$ and adds into $S$. Then, the algorithm invokes Algorithm~\ref{alg:update} (Update($I[1][1:8], D[1][1:8], v_2, 1$)) to update $D[1][1:8]$. After this step, we can obtain that only $D[1][2]$, $D[1][1]$, $D[1][3]$, and $D[1][5]$ need to be updated, and they are re-set to 0, 1, 1, and 1 respectively. Similar arguments can be used for analyzing the second round. Here we only report the result, and omit the details for brevity. In the second round, the algorithm adds $v_7$ into the answer set. Therefore, the algorithm outputs $\{v_2, v_7\}$ as the targeted nodes.
\end{example}

We analyze the time and space complexity of Algorithm~\ref{alg:approxgreedy} as follows. First, to build the inverted index (line~1), Algorithm~\ref{alg:indexcons} takes $O(RLn)$ time complexity. Second, to estimate the marginal gain for every node, the algorithm needs to invoke Algorithm~\ref{alg:approxgain} $O(n)$ times. We can derive that the time complexity of this step (line~5) is $O(nRL)$, because the algorithm only needs to access the entire inverted index once and the size of the inverted index is bounded by $O(nRL)$. Third, to update $D[1:R][1:n]$, Algorithm~\ref{alg:update} takes at most $O(Rn)$ time. Put it all together, the time complexity of Algorithm~\ref{alg:approxgreedy} is $O(kRLn)$, which is linear w.r.t.\ the graph size ($R$, $k$, and $L$ are small constants). For the space complexity, the algorithm needs to maintain two arrays: the inverted index $I[1:R][1:n]$ and the array $D[1:R][1:n]$. Clearly, $I[1:R][1:n]$ and $D[1:R][1:n]$ are bounded by $O(RLn)$ and $O(Rn)$ respectively. Therefore, the space complexity of Algorithm~\ref{alg:approxgreedy} is $O(nRL+m)$.

Note that in Algorithm~\ref{alg:approxgreedy}, each marginal gain is estimated by the same $R$ $L$-length random walks. Since the $L$-length random walks are independent of one another, the estimator is able to achieve high accuracy. As a result, the approximation factor of  Algorithm~\ref{alg:approxgreedy} is $1-1/e-\epsilon$ by setting an appropriate $R$. In the experiments, we find that the effectiveness of Algorithm~\ref{alg:approxgreedy} are comparable with the DP-based greedy algorithm even when $R$ is a small value (e.g., $R=100$).

\section{Experiments}
\label{sec:experiments}
In this section, we conduct extensive experiments over both synthetic and real-world graphs. We aim at evaluating the effectiveness, efficiency and scalability of our algorithms. In the following, we first describe the experimental setup and then report the results.

\subsection{Experimental setup}
\label{subsec:expsetup}
\stitle{Different algorithms}: Since the proposed random-walk domination problems are novel, we are not aware of any algorithm that addresses to these problems in the literature. Intuitively, the high-degree nodes are more easily reached by the other nodes. Therefore, to maximize the expected number of reached nodes, a reasonable baseline algorithm is to select the top-$k$ high-degree nodes as the targeted nodes. For convenience, we refer to this baseline algorithm as the  \emph{Degree} algorithm. The second baseline is the traditional dominating-set-based algorithm \cite{98dominate1}. A dominating set is a subset of nodes $D \subset V$ such that every node in $V$ is either in $D$ or a neighbor of some nodes in $D$ \cite{98dominate1}. By this definition, every node can only dominate its neighbors. In our problems, since we have a cardinality constraint, i.e, $|S| \le k$, we cannot select the entire dominating set. Instead, we turns to select $k$ nodes such that they can dominate as many nodes as possible. Note that here the concept of domination is based on the definition of traditional dominating set. Specifically, let $S$ be the set of targeted nodes. Initially, $S$ is an empty set. The algorithm works in $k$ rounds. In each round, the algorithm selects a node $v$ such that $v = \arg \mathop {\max }\nolimits_{u \in V/S} |N(\{ u\} ) - N(S)|$, where $N(S)$ denotes the set of immediate neighbors of nodes in $S$. Then, the algorithm adds $v$ into the set $S$, and goes to the next round. We call this algorithm the \emph{Dominate} algorithm.

We compare two proposed algorithms with the above two baseline algorithms. The first algorithm is the DP-based greedy algorithm, in which the marginal gain is calculated by the DP algorithm. The second algorithm is the approximate greedy algorithm i.e., Algorithm~\ref{alg:approxgreedy}. Both of them are used to solve both problem (1) (Eq.~(\ref{eq:problem1r})) and problem (2) (Eq.~(\ref{eq:problem2})). Here we do not report the results of the sampling-based greedy algorithm because the approximate greedy algorithm is more efficient than such an algorithm. For convenience, we refer to the first algorithm for solving problem (1) and problem (2) as \emph{DPF1} and \emph{DPF2} respectively. Similarly, we call the second algorithm for solving problem (1) and problem (2) as \emph{ApproxF1} and \emph{ApproxF2} respectively.

\stitle{Evaluation metrics: }Two metrics are used to evaluate the effectiveness of different algorithms. The first metric is the average hitting time which is defined as $M_1 (S) = \sum\nolimits_{u \in V\backslash S} {h_{uS}^L } / |V \backslash S| $, where $S$ denotes the set of selected nodes by a algorithm. This metric inversely measures the effectiveness of the algorithm. In other words, the smaller the $M_1 (S)$ is, the more effective the algorithm is. 
The second metric is the expected number of nodes that hit a node in $S$ via an $L$-length random walk. The formula of the second metric is given by $M_2 (S) = \sum\nolimits_{u \in V} {\mathbb{E}[X_{uS}^L ]} $. The larger $M_2(S)$ implies the higher effectiveness of the algorithm. 
For convenience, we refer to the first metric and the second metric as \emph{AHT} and \emph{EHN} respectively. Note that to compute these metrics, we uses the sampling algorithm described in Algorithm~\ref{alg:sampling} and set the sample size $R=500$. To evaluate the efficiency of different algorithms, we record the running time, which is measured by the wall-clock time.

\begin{table}[t]
\begin{center}
\caption[]{Summary of the datasets} \label{tbl:data}
\begin{tabular}{l|c|c}
\hline
Name & \# of nodes & \# of edges  \\
\hline \hline
CAGrQc & 5,242  & 28,968  \\
CAHepPh & 12,008 & 236,978  \\
Brightkite & 58,228 & 428,156  \\
Epinions & 75,872 & 396,026   \\
\hline
\end{tabular}\vspace*{-0.5cm}
\end{center}
\end{table}

\stitle{Datasets: }We use four real-world datasets in our experiments: \emph{CAGrQc}, \emph{CAHepPh}, \emph{Brightkite}, and \emph{Epinions}. The \emph{CAGrQc} and \emph{CAHepPh} datasets are co-authorship networks which represent the co-authorship over two different areas in physics respectively. The \emph{Brightkite} is a location-based social network dataset, where the users in Brightkite can check-in spots and share their location information with their friends. The \emph{Epinions} is a trust social network dataset, where the edge represents the trust relationship between two users. All the four datasets are downloaded from Stanford network data collections \cite{standforddata}. The detailed statistic information of the datasets are shown in Table~\ref{tbl:data}.

\stitle{Experimental environment: }We conduct all the experiments on a Windows XP PC with 2xQuad-Core Intel Xeon 2.66 GHz CPU, and 8GB memory. All the algorithms are implemented in C++.

\subsection{Experimental Results}
\label{subsec:expresults}

\begin{figure}[t]
\begin{center}
\includegraphics[width=\columnwidth, height=5cm]{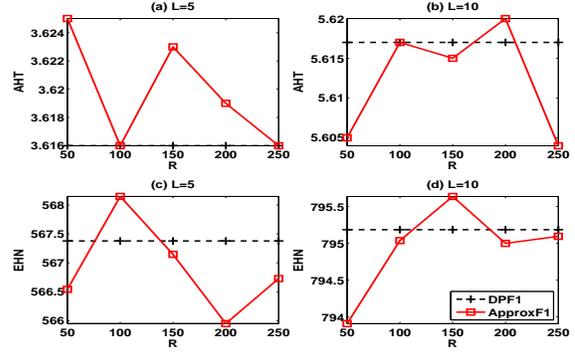}
\end{center}\vspace*{-2em}
\caption[]{Comparison of effectiveness of \emph{DPF1} and  \emph{ApproxF1}}
\label{fig:accuratef1} \vspace*{-0.3cm}
\end{figure}

\begin{figure}[t]
\begin{center}
\includegraphics[width=\columnwidth, height=5cm]{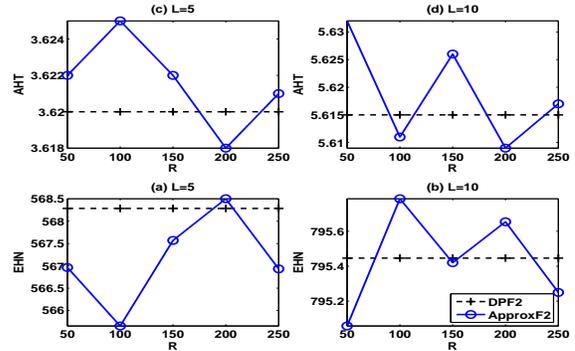}
\end{center}\vspace*{-2em}
\caption[]{Comparison of effectiveness of \emph{DPF2} and  \emph{ApproxF2}}
\label{fig:accuratef2} \vspace*{-0.3cm}
\end{figure}

\stitle{Performance of the approximate greedy algorithms: }Here we compare the effectiveness and efficiency of the approximate greedy algorithms (\emph{ApproxF1} and \emph{ApproxF2}) with those of the DP-based greedy algorithm (\emph{DPF1} and \emph{DPF2}). Due to the expensive time and space complexity of the \emph{DPF1} and \emph{DPF2} algorithms, these two algorithms can only work well on very small datasets. To this end, we generate a small synthetic graph with 1000 nodes and 9956 edges based on a commonly-used power-law random graph model \cite{99sfnet}. We set the parameter $k$ to 30 which denotes the number of selected nodes of different algorithms, and set the parameter $L$ in the $L$-length random walk model to $5$ and $10$ respectively. Similar results can be observed for other values of $k$ and $L$. The results are shown in Fig.~\ref{fig:accuratef1} and Fig.~\ref{fig:accuratef2}. Specifically, Fig.~\ref{fig:accuratef1} depicts the comparison of effectiveness of \emph{DPF1} and \emph{ApproxF1} algorithms. The black dash line in Fig.~\ref{fig:accuratef1} describes the effectiveness of the \emph{DPF1} algorithm, while the red solid curve depicts the effectiveness of the \emph{ApproxF1} algorithm as a function of the parameter $R$, denoting the number of samples used to estimate the marginal gain. As can be seen in Fig.~\ref{fig:accuratef1}, the \emph{ApproxF1} algorithm is very accurate when the number of samples is greater than or equal to 50. For example, in Fig.~\ref{fig:accuratef1}(a), the greatest difference of \emph{AHT} between \emph{DPF1} and \emph{ApproxF1} algorithms is around 0.01, which is achieved at $R=50$. Moreover, when $R=100$, the result of the \emph{ApproxF1} algorithm matches the result of the \emph{DPF1} algorithm. In Fig.~\ref{fig:accuratef1}(c), we can see that the expected number nodes that can hit the selected nodes calculated by the \emph{ApproxF1} algorithm is very close to the expected number of nodes computed by the \emph{DPF1} algorithm. The maximal difference of \emph{EHN} between \emph{DPF1} and \emph{ApproxF1} algorithms is around 1.5, which is achieved at $R=200$.

Fig.~\ref{fig:accuratef2} illustrates the comparison of effectiveness of \emph{DPF2} and \emph{ApproxF2} algorithms. Similarly, from Fig.~\ref{fig:accuratef2}, we can observe that the effectiveness of the \emph{ApproxF2} algorithm is very close to that of the \emph{DPF2} algorithm. In Fig.~\ref{fig:accuratef2}(a), for instance, the maximal difference of \emph{AHT} between the \emph{DPF2} and \emph{ApproxF2} algorithms is smaller than 0.01 (obtained at $R=100$). Hence, for both \emph{AHT} and \emph{EHN} metrics, the approximate greedy algorithms work very well with a small $R$ value. These results are consistent with the theoretical analysis in Section~\ref{subsec:approxgreedy}.

\begin{figure}[t]
\begin{center}
\includegraphics[width=\columnwidth, height=2.3cm]{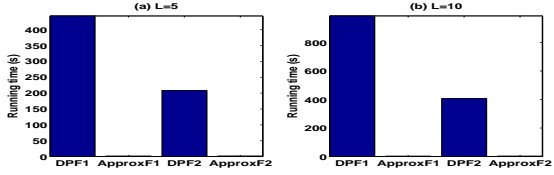}
\end{center}\vspace*{-2em}
\caption[]{Comparison of running time: DP-based greedy algorithms vs approximate greedy algorithms.}
\label{fig:timedpvsapprox} \vspace*{-0.3cm}
\end{figure}

Now we compare the running time of the approximate greedy algorithms (\emph{ApproxF1} and \emph{ApproxF2}) with that of the DP-based greedy algorithms (\emph{DPF1} and \emph{DPF2}). The results are reported in Fig.~\ref{fig:timedpvsapprox}. From Fig.~\ref{fig:timedpvsapprox}, we can clearly see that the running time of the \emph{DPF1} and \emph{DPF2} algorithms are significantly longer than the running time of the \emph{ApproxF1} and \emph{ApproxF2} algorithms, where the running time of the \emph{ApproxF1} and \emph{ApproxF2} algorithms are recorded at $R=250$. For example, in Fig.~\ref{fig:timedpvsapprox}(a), the running time of the \emph{DPF1} algorithm is larger than 400 seconds, while the running time of the \emph{ApproxF1} algorithm is around 2 seconds. That is to say, the efficiency of the \emph{ApproxF1} algorithm is better than that of the \emph{DPF1} algorithm by 200 times. It is worth mentioning that the running time of the \emph{DPF1} is twice as much as the running time of the \emph{DPF2}. This is because the \emph{DPF1} algorithm needs an extra ``addition operation'' for computing the hitting time (Eq.~(\ref{eq:ghitrur})) comparing with the \emph{DPF2} algorithm. In addition, the running time of different algorithms when $L=10$ is twice as much as the running time of different algorithms when $L=5$.

\begin{figure}[t]
\begin{center}
\includegraphics[width=\columnwidth, height=2.3cm]{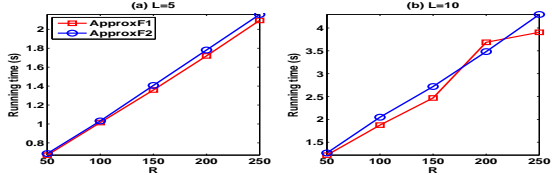}
\end{center}\vspace*{-2em}
\caption[]{Running time as a function of $R$}
\label{fig:timeapprox} \vspace*{-0.3cm}
\end{figure}

We also study the running time of the \emph{ApproxF1} and \emph{ApproxF2} algorithms as a function of the parameter $R$. The results are shown in Fig.~\ref{fig:timeapprox}. As observed, the running time of the \emph{ApproxF1} and \emph{ApproxF2} algorithms increase linearly as $R$ increases, which conforms with that the time complexity of the approximate greedy algorithms is linear w.r.t.\ $R$.

\begin{figure}[t]
\begin{center}
\includegraphics[width=\columnwidth, height=5cm]{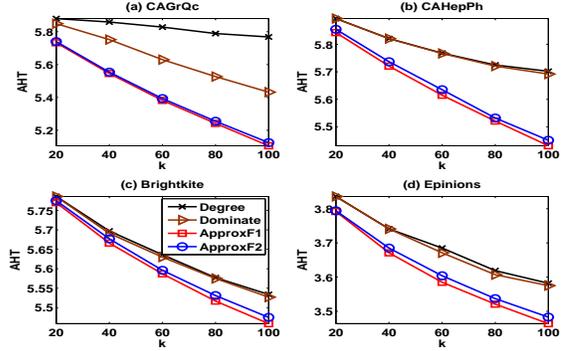}
\end{center}\vspace*{-2em}
\caption[]{Comparison of \emph{AHT} of different algorithms}
\label{fig:aht} \vspace*{-0.3cm}
\end{figure}

\begin{figure}[t]
\begin{center}
\includegraphics[width=\columnwidth, height=5cm]{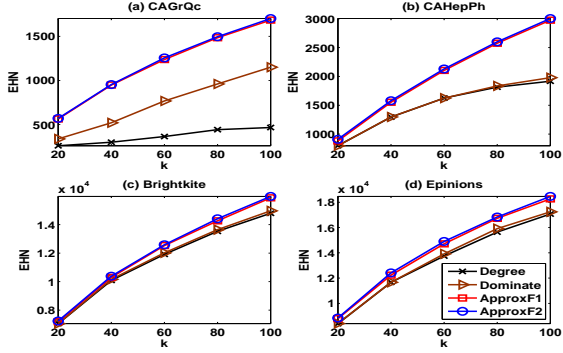}
\end{center}\vspace*{-2em}
\caption[]{Comparison of \emph{EHN} of different algorithms}
\label{fig:ehn} \vspace*{-0.3cm}
\end{figure}

\stitle{Effectiveness of different algorithms: }Here we compare the effectiveness of different algorithms over four real-world datasets. As indicated in the previous experiment, under both \emph{AHT} and \emph{EHN} metrics, there is no significant difference between the \emph{ApproxF1} (\emph{ApproxF2}) algorithm and the \emph{DPF1} (\emph{DPF2}) algorithm. Furthermore, the former algorithms are more efficient than the latter algorithms up to two orders of magnitude. Hence, in the following experiments, for the greedy algorithms, we only report the results obtained by the \emph{ApproxF1} and \emph{ApproxF2} algorithms. For these algorithms, we set the parameter $R$ to 100 in all the following experiments without any specific statements, because $R=100$ is enough to ensure good accuracy as shown in the previous experiment. For all the algorithms, we set the parameter $L$ to 6, and similar results can be observed for other $L$ values. Fig.~\ref{fig:aht} and Fig.~\ref{fig:ehn} describe the results of different algorithms over four real-world datasets under \emph{AHT} and \emph{EHN} metrics respectively. From Fig.~\ref{fig:aht}, we can see that both the \emph{ApproxF1} and \emph{ApproxF2} algorithms are significantly better than the two baselines in all the datasets used. As desired, for all the algorithms, the \emph{AHT} decreases as $k$ increases. In addition, we can see that the \emph{ApproxF1} algorithm slightly outperforms the \emph{ApproxF2} algorithms, because the \emph{ApproxF1} algorithm directly optimizes the \emph{AHT} metric. Also, we can observe that the \emph{Dominate} algorithm is slightly better than the \emph{Degree} algorithm in \emph{CAHepph}, \emph{Brightkite}, and \emph{Epinions} datasets. In CAGrQc datasets, however, the \emph{Degree} algorithm performs poorly, and the \emph{Dominate} algorithm significantly outperforms the \emph{Degree} algorithm. Similarly, as can be seen in Fig.~\ref{fig:ehn}, the \emph{ApproxF1} and \emph{ApproxF2} algorithms substantially outperform the baselines over all the datasets under the \emph{EHN} metric. Moreover, we can see that the \emph{ApproxF2} algorithm is slightly better than the \emph{ApproxF1} algorithm, because the \emph{ApproxF2} algorithm directly maximizes the \emph{EHN} metric. Note that, under both \emph{AHT} and \emph{EHN} metrics, the gap between the curves of the approximate greedy algorithms and those of the two baselines increases with increasing $k$. The rationale is that the approximate greedy algorithms are near-optimal which achieve $1-1/e-\epsilon$ approximation factor, and such approximation factor is independent of the parameter $k$. The two baselines, however, are without any performance guarantee, thus the effectiveness of these two algorithms would decrease as $k$ increases. These results are consistent with our theoretical analysis in Section~\ref{sec:algs}.

\begin{figure}[t]
\begin{center}
\includegraphics[width=\columnwidth, height=2.5cm]{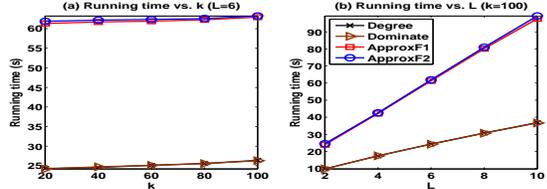}
\end{center}\vspace*{-2em}
\caption[]{Comparison of running time of different algorithms in Epinonios dataset}
\label{fig:timecomparison} \vspace*{-0.3cm}
\end{figure}

\stitle{Efficiency of different algorithms: }Here we evaluate the efficiency of different algorithms. Fig.~\ref{fig:timecomparison} shows the comparison of the running time of different algorithms over the Epinions dataset. Similar results can be obtained in other datasets. In particular, Fig.~\ref{fig:timecomparison}(a) depicts the running time of different algorithms as a function of the parameter $k$. Here the parameter $L$ is set to 6. In particular, from Fig.~\ref{fig:timecomparison}(a), we are able to observe that the running time of the \emph{ApproxF1} and \emph{ApproxF2} algorithms are around 2.5 times longer than the running time of the \emph{Degree} and \emph{Dominate} algorithms. Fig.~\ref{fig:timecomparison}(b) illustrates the running time of different algorithms as a function of the parameter $L$, where we set the parameter $k$ to 100. As can be observed in Fig.~\ref{fig:timecomparison}(b), the running time of the \emph{ApproxF1} and \emph{ApproxF2} algorithms are longer than that of the \emph{Degree} and \emph{Dominate} algorithms by 2.7 times at most. For example, when $L=10$, the running time of the \emph{ApproxF1} is 99 seconds, while the running time of the \emph{Degree} algorithm is 37 seconds. These results indicate that the approximate greedy algorithms is only a small constant times longer than that of the \emph{Degree} algorithm, which are consistent with the complexity analysis in Section~\ref{subsec:approxgreedy}.

\begin{figure}[t]
\begin{center}
\includegraphics[width=\columnwidth, height=2.5cm]{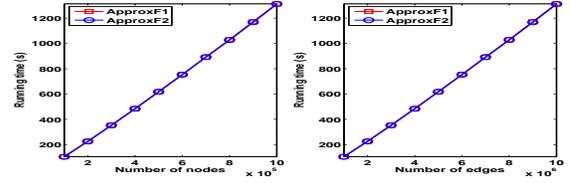}
\end{center}\vspace*{-2em} \caption[]{Scalability testing}
\label{fig:scalability} \vspace*{-0.3cm}
\end{figure}

\stitle{Scalability testing: }Here we evaluate the scalability of the \emph{ApproxF1} and \emph{ApproxF2} algorithms. To this end, we generate ten large synthetic graphs according to a widely-used power-law random graph model \cite{99sfnet}. More specifically, we generate ten graphs $G_1, \cdots, G_{10}$ such that $G_i$ has $i \times 0.1$ million nodes and $i$ million edges for $i=1, \cdots, 10$. Fig.~\ref{fig:scalability} shows the results of the \emph{ApproxF1} and \emph{ApproxF2} algorithms w.r.t.\ the number of nodes (left panel) and w.r.t.\ the number of edges (right panel). Here we set the parameter $L=6$ and $k=100$. Similar results can be observed for other values of $L$ and $k$. From Fig.~\ref{fig:scalability}, we find that both the \emph{ApproxF1} and \emph{ApproxF2} algorithms scale linearly w.r.t.\ both the number of nodes and the number of edges, which is consistent with the linear time complexity (w.r.t.\ the graph size) of the algorithm.

\begin{figure}[t]
\begin{center}
\includegraphics[width=\columnwidth, height=5cm]{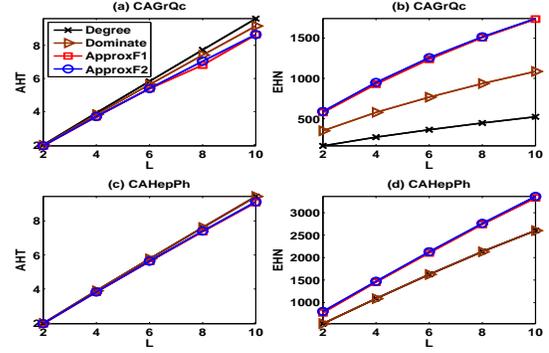}
\end{center}\vspace*{-2em}
\caption[]{Effect of parameter $L$}
\label{fig:effectl} \vspace*{-0.3cm}
\end{figure}

\stitle{Effect of parameter $L$: }Here we study the effect of parameter $L$. Fig.~\ref{fig:effectl} reports the results in \emph{CAGrQc} and \emph{CAHepPh} datasets given $k=60$. Similar results can be observed in other datasets and other values of $k$ as well. From Fig.~\ref{fig:effectl}(a-d), we can see that both the \emph{AHT} and \emph{EHN} by different algorithms increase as $L$ increases. Recall that the hitting time is bounded by $L$, and the hitting time of a node that cannot hit the targeted nodes is set to $L$. Therefore, the average hitting time will increase if $L$ increase. Clearly, with $L$ increasing, the number of nodes that can hit the targeted nodes will increase, thereby the \emph{EHN} of different algorithms will increase. In addition, we find that the gap between the curves of the  \emph{ApproxF1} and \emph{ApproxF2} algorithms and the curves of the baselines increases as $L$ increases, which suggests that the \emph{ApproxF1} and \emph{ApproxF2} algorithms perform very well for large $L$ values. 

\section{Conclusions}
\label{sec:concl}

In this paper, we introduce and formulate two random-walk domination problems in graphs motivated by a number of applications such as the item placement in social networks, the resource placement in P2P network, and the advertisements placement in advertisement networks. We show that these two problems are an instance of submodular set function maximization with cardinality constraint problem. Based on this, we propose a dynamic programming (DP) based greedy algorithm with $1-1/e$ approximation factor to solve them. The DP-based greedy algorithm, however, is not very efficient because of the expensive marginal gain evaluation. To further accelerate the greedy algorithm, we present an approximate greedy algorithm with liner time complexity w.r.t.\ the graph size. We show that the approximate greedy algorithm is also with near-optimal performance guarantee. Extensive experiments are conducted to evaluate the proposed algorithms. The results demonstrate the effectiveness, efficiency, and scalability of the proposed algorithms.

There are a number of future directions needed to further investigation. First, since the objective functions of Problem (1) and Problem (2) are submodular, one may combines these two objective functions (e.g., by a positive weights, it is still submodular) and study the problem of optimizing both the total hitting time and the expected number of nodes that hit the targeted set simultaneously. Second, Problem (2) is to count the expected number of nodes that are dominated by the targeted set. It would be interesting to extend this problem to count the expected number of edges that are traversed by the $L$-length random walk starting from any node to the targeted set. Finally, Problem (2) is to maximize the expected number of nodes. A complementary problem is that given a parameter $\alpha \in [0, 1]$, the goal is to find the minimum number of targeted nodes such that they can dominate at least $\alpha n$ number of nodes in expectation. It would also be interesting to devise efficient algorithms for this problem.

\section*{Appendix} 
\stitle{Proof of Theorem~\ref{thm:trunhitting}}:
By definition, we have the following facts.

\stitle{Fact 1: } If {  $0 < i < L$}, we have {  $\Pr [T_{uv}^L = i]
= \sum\nolimits_{w \in V} {p_{uw} \Pr [T_{wv}^L = i - 1]}, $} and
if {  $i = L$}, we have {  $\Pr [T_{uv}^L  = i] = \sum\nolimits_{w \in V}
{p_{uw} \Pr [T_{wv}^L  \ge i - 1]}$}.

\stitle{Fact 2: }If {  $0 < i < L - 1$}, we have
{  $\Pr[T_{uv}^{L-1} = i] = \Pr [T_{uv}^L = i]$},
and if {  $i = L - 1$}, we have {  $\Pr [T_{uv}^{L-1} = i] = \Pr [T_{uv}^L = i] + \Pr [T_{uv}^L = L]$}.

Equipped with the above two facts, we can prove the theorem as
follows. Clearly, if $u=v$, then $T^L_{uv}=0$, and thereby
$h^L_{uv}=0$. If $u \ne v$, by definition, we have
\begin{equation}
\begin{array}{l}
 h_{uv}^L  = \mathbb{E}[T_{uv}^L ] = \sum\nolimits_{i = 1}^L {i\Pr [T_{uv}^L  = i]}  \\
  \quad \quad \quad \quad \quad \quad \; = \sum\nolimits_{i = 1}^{L - 1} {i\Pr [T_{uv}^L  = i]}  + L\Pr [T_{uv}^L  = L] \\
 \quad \quad \quad \quad \quad \quad \; = \sum\nolimits_{i = 1}^{L - 1} {i\sum\nolimits_{w \in V} {p_{uw} \Pr [T_{wv}^L  = i - 1]} }  \\
 \quad \quad \quad \quad \quad \quad \; \quad \quad + L\sum\nolimits_{w \in V} {p_{uw} \Pr [T_{wv}^L  \ge L - 1]}  \\
 \quad \quad \quad \quad \quad \quad \; = \sum\nolimits_{i = 1}^L {i\sum\nolimits_{w \in V} {p_{uw} \Pr [T_{wv}^L  = i - 1]} } \\
 \quad \quad \quad \quad \quad \quad \; \quad \quad  + L\sum\nolimits_{w \in V} {p_{uw} \Pr [T_{wv}^L  = L]},  \\
 \end{array}
 \label{eq:proofeq1}
\end{equation}
where the third equation holds due to Fact 1. Then, we can further
reduce Eq.~(\ref{eq:proofeq1}) as follows.
\begin{equation}
  \label{eq:proofeq2}
  \begin{array}{l}
 h_{uv}^L  = \sum\nolimits_{i = 1}^L {(i - 1)\sum\nolimits_{w \in V} {p_{uw} \Pr [T_{wv}^L  = i - 1]} }    \\
 \quad \quad \quad \quad \quad + \sum\nolimits_{i = 1}^L {\sum\nolimits_{w \in V} {p_{uw} \Pr [T_{wv}^L  = i - 1]} }    \\
 \quad \quad \quad \quad \quad + \sum\nolimits_{w \in V} {p_{uw} \Pr [T_{wv}^L  = L]} \\
 \quad \quad \quad \quad \quad + (L - 1)\sum\nolimits_{w \in V} {p_{uw} \Pr [T_{wv}^L  = L]}  \\
 \quad \quad  = \sum\nolimits_{i = 1}^L {(i - 1)\sum\nolimits_{w \in V} {p_{uw} \Pr [T_{wv}^L  = i - 1]} }    \\
 \quad \quad \quad \quad \quad + (L - 1)\sum\nolimits_{w \in V} {p_{uw} \Pr [T_{wv}^L  = L]}  + 1, \\
 \end{array}
\end{equation}
where the equality holds is owing to $ \sum\nolimits_{i = 1}^L {\Pr
[T_{wv}^L = i]} = 1$ and $ \sum\nolimits_{w \in V} {p_{uw} }  = 1$.
Based on Eq.~(\ref{eq:proofeq2}) and Fact 2, we have
\begin{equation}
\begin{array}{l}
 h_{uv}^L  = \sum\nolimits_{i = 1}^{L - 1} {i\sum\nolimits_{w \in V} {p_{uw} \Pr [T_{wv}^L  = i]} }  \\
 \quad  + (L - 1)\sum\nolimits_{w \in V} {p_{uw} \Pr [T_{wv}^L  = L]}  + 1 \\
  = \sum\nolimits_{i = 1}^{L - 2} {i\sum\nolimits_{w \in V} {p_{uw} \Pr [T_{wv}^L  = i]} }  \\
 \quad + (L - 1)\sum\nolimits_{w \in V} {p_{uw} (\Pr [T_{wv}^L  = L - 1] + \Pr [T_{wv}^L  = L]} ) + 1 \\
  = \sum\nolimits_{i = 1}^{L - 2} {i\sum\nolimits_{w \in V} {p_{uw} \Pr [T_{wv}^{L - 1}  = i]} }  \\
 \quad + (L - 1)\sum\nolimits_{w \in V} {p_{uw} (\Pr [T_{wv}^{L - 1}  = L - 1]} ) + 1 \quad \{$By Fact 2$\} \\
  = \sum\nolimits_{i = 1}^{L - 1} {i\sum\nolimits_{w \in V} {p_{uw} \Pr [T_{wv}^{L - 1}  = i]} }  + 1 \\
  = 1 + \sum\nolimits_{w \in V} {p_{uw} h_{wv}^{L - 1} }.  \\
 \end{array}\label{eq:proofeq3}
\end{equation}
This completes the proof.

\stitle{Proof of Theorem~\ref{thm:f1submodular}}:
  First, it is easy to check that $F_1(\emptyset) = 0$. Second, we prove that $F_1(S)$ is a non-increasing set function. Let $S \subseteq T \subseteq V$ be two subsets of $V$. Then, for any node $u \in V \backslash T$, we claim that
  \begin{equation}   \label{eq:inequalityhit}
  h_{uT}^L  \le h_{uS}^L.
  \end{equation}
  We shall prove the above inequality by induction. By definition, we have $h_{uT}^0  = h_{uS}^0  = 0$ and $h_{uT}^1  = h_{uS}^1  = 1$. Therefore, the inequality defined in Eq.~(\ref{eq:inequalityhit}) holds if $L=0$ and $L=1$. Suppose that $h_{uT}^L  \le h_{uS}^L$ holds given $L=\alpha > 1$. Below, we show that the inequality still holds if $L=\alpha + 1$. By Eq.~(\ref{eq:ghitrur}), we have
    \[
\begin{array}{l}
 h_{uS}^{\alpha  + 1}  = 1 + \sum\nolimits_{w \notin S} {p_{uw} h_{wS}^\alpha  } \\
   \quad \;\;\;\; \;  = 1 + \sum\nolimits_{w \notin T} {p_{uw} h_{wS}^\alpha  }  + \sum\nolimits_{w \in T\backslash S} {p_{uw} h_{wS}^\alpha  }  \\
 \quad \;\;\;\; \; \ge 1 + \sum\nolimits_{w \notin T} {p_{uw} h_{wS}^\alpha  }  \ge 1 + \sum\nolimits_{w \notin T} {p_{uw} h_{wT}^\alpha  }   = h_{uT}^{\alpha  + 1},
 \end{array}
\]
 where the last inequality holds due to the induction assumption. Based on Eq.~(\ref{eq:inequalityhit}), we have
  \[
\begin{array}{l}
 F_1 (S) - F_1 (T) = \sum\nolimits_{u \in V\backslash T} {h_{uT}^L }  - \sum\nolimits_{u \in V\backslash S} {h_{uS}^L }  \\
  \quad \quad \quad \quad \quad \quad \;\;\;\le \sum\nolimits_{u \in V\backslash T} {(h_{uT}^L }  - h_{uS}^L ) \le 0.
 \end{array}
\]
Thus, $F_1(S)$ is a non-increasing set function as desired. Finally, we prove the submodularity property of $F_1(S)$. Let $T_u=T \cup \{u\}$ and $S_u=S \cup \{u\}$. Let $\sigma _u (S)=F_1(S_u)-F_1(S)$ be the marginal gain. Then, we have { \[\sigma _u (S) = \sum\nolimits_{w \in V\backslash S} {h_{wS}^L }  - \sum\nolimits_{w \in V\backslash S_u } {h_{wS_u }^L }\] and { \[\sigma _u (T) = \sum\nolimits_{w \in V\backslash T} {h_{wT}^L }  - \sum\nolimits_{w \in V\backslash T_u } {h_{wT_u }^L }.\] To prove the submodularity of $F_1(S)$, we show $\sigma _u (T) \le \sigma _u (S)$ as follows:
\begin{equation}  \label{eq:submodularf1ieq}
\begin{array}{l}
 \sigma _u (S) - \sigma _u (T) \\
 \quad \quad \quad  = (\sum\nolimits_{w \in V\backslash S} {h_{wS}^L }  - \sum\nolimits_{w \in V\backslash T} {h_{wT}^L } ) \\
 \quad \quad \quad \quad \quad  \quad \quad -(\sum\nolimits_{w \in V\backslash S_u } {h_{wS_u }^L } - \sum\nolimits_{w \in V\backslash T_u } {h_{wT_u }^L } ) \\
 \quad \quad \quad   = \sum\nolimits_{w \in T\backslash S} {(h_{wS}^L  - h_{wT}^L )}  - \sum\nolimits_{w \in T\backslash S} {(h_{wS_u }^L  - } h_{wT_u }^L ) \\
 \quad \quad \quad  = \sum\nolimits_{w \in T\backslash S} {(h_{wS}^L  - h_{wS_u }^L )} \ge 0. \\
 \end{array}
\end{equation}
 Since $\sum\nolimits_{w \in T\backslash S} {h_{wT}^L }=0$ and $\sum\nolimits_{w \in T\backslash S } {h_{wT_u }^L }=0$ by Eq.~(\ref{eq:ghitrur}), the third equality of the above equation holds. To prove the last inequality of Eq.~(\ref{eq:submodularf1ieq}), we can use a similar induction argument which is applied to prove Eq.~(\ref{eq:inequalityhit}). We omit the details for brevity. Put it all together, we conclude that $F_1(S)$ is a non-increasing submodular set function with $F_1(\emptyset)=0$. Therefore, the theorem is established.

\stitle{Proof of Theorem~\ref{thm:f2submodular}}:
  First, by definition, $X_{uS} ^L$ equals to zero if $S=\emptyset$, which results in $F_2(\emptyset)
= 0$. Second, we show the non-increasing property of $F_2(S)$. Let $S \subseteq T \subseteq V$ be two subsets of $V$. By the linearity of expectation, we have $F_2 (S) = \sum\nolimits_{w \in V} {\mathbb{E}(X_{wS}^L )}=\sum\nolimits_{w \in V} {p_{wS}^L }$. Let $p_{wv}^L$ be the probability of that $w$ hits $v$ by an $L$-length random walk. Then, we have $p_{wS}^L  = 1 - \prod\nolimits_{v \in S} {(1 - p_{wv}^L )}$. Further, we have
\[
\begin{array}{l}
 F_2 (S) - F_2 (T) = \sum\nolimits_{w \in V} {(p_{wS}^L  - } p_{wT}^L ) \\
  = \sum\nolimits_{w \in V} {((1 - \prod\nolimits_{v \in S} {(1 - p_{wv}^L )} ) - } (1 - \prod\nolimits_{v \in T} {(1 - p_{wv}^L )} )) \\
  = \sum\nolimits_{w \in V} ( \prod\nolimits_{v \in T} {(1 - p_{wv}^L )}  - \prod\nolimits_{v \in S} {(1 - p_{wv}^L )} )
  \le 0. \\
 \end{array}
\]
Therefore, $F_2(S)$ is a non-increasing set function. Finally, we prove that $F_2(S)$ is a submodular set function. Let $u \in V \backslash T$, $S_u = S \cup \{u\}$, and $T_u = T \cup \{u\}$. Further, we let $\rho_u(S)=F_2(S \cup \{u\})-F_2(S)$ be the marginal gain. Then, we have 
\[
\rho _u (S) = \sum\nolimits_{w \in V} {(\prod\nolimits_{v \in S} {(1 - p_{wv}^L )}  - \prod\nolimits_{v \in S_u } {(1 - p_{wv}^L )} )}
\]
and
\[
\rho _u (T) = \sum\nolimits_{w \in V} {(\prod\nolimits_{v \in T} {(1 - p_{wv}^L )}  - \prod\nolimits_{v \in T_u } {(1 - p_{wv}^L )} )}.
\]
In the following, we show that $\rho _u (S) \ge \rho _u (T)$. Specifically, we have
\[
\begin{array}{l}
 \rho _u (S) - \rho _u (T) \\
 \quad \quad = \sum\nolimits_{w \in V} {((\prod\nolimits_{v \in S} {(1 - p_{wv}^L )}  - \prod\nolimits_{v \in T} {(1 - p_{wv}^L )} )}  \\
   \quad \quad  \quad \quad  \quad \quad - (\prod\nolimits_{v \in S_u } {(1 - p_{wv}^L )}  - \prod\nolimits_{v \in T_u } {(1 - p_{wv}^L )} )) \\
   \quad \quad = \sum\nolimits_{w \in V} {((1 - \prod\nolimits_{v \in T\backslash S} {(1 - p_{wv}^L )} )\prod\nolimits_{v \in S} {(1 - p_{wv}^L )} }  \\
 \quad \quad  \quad \quad  \quad \quad - (1 - \prod\nolimits_{v \in T\backslash S} {(1 - p_{wv}^L )} )\prod\nolimits_{v \in S_u } {(1 - p_{wv}^L )}  \\
  \quad \quad = \sum\nolimits_{w \in V} {((1 - \prod\nolimits_{v \in T\backslash S} {(1 - p_{wv}^L )} )p_{wu}^L \prod\nolimits_{v \in S} {(1 - p_{wv}^L )} }  \\
  \quad \quad \ge 0. \\
 \end{array}
\]
This completes the proof.

{
\bibliographystyle{abbrv}
\bibliography{rwcover}

\begin{thebibliography}{10}

\bibitem{99sfnet}
A.-L. Barabasi and R.~Albert.
\newblock Emergence of scaling in random networks.
\newblock {\em science}, 1999.

\bibitem{08adhocdominate}
M.~Couture, M.~Barbeau, P.~Bose, and E.~Kranakis.
\newblock Incremental construction of k-dominating sets in wireless sensor
  networks.
\newblock {\em Ad Hoc {\&} Sensor Wireless Networks}, 5(1-2):47--68, 2008.

\bibitem{12pvldbfilterplacement}
D.~Erd{\"o}s, V.~Ishakian, A.~Lapets, E.~Terzi, and A.~Bestavros.
\newblock The filter-placement problem and its application to minimizing
  information multiplicity.
\newblock {\em PVLDB}, 5(5):418--429, 2012.

\bibitem{98jacmsetcover}
U.~Feige.
\newblock A threshold of ln {\it n} for approximating set cover.
\newblock {\em J. ACM}, 45(4), 1998.

\bibitem{06p2psearch}
C.~Gkantsidis, M.~Mihail, and A.~Saberi.
\newblock Random walks in peer-to-peer networks: Algorithms and evaluation.
\newblock {\em Perform. Eval.}, 63(3):241--263, 2006.

\bibitem{98connecteddominateset}
S.~Guha and S.~Khuller.
\newblock Approximation algorithms for connected dominating sets.
\newblock {\em Algorithmica}, 20(4):374--387, 1998.

\bibitem{98dominate2}
T.~W. Haynes, S.~T. Hedetniemi, and P.~J. Slater.
\newblock {\em Domination in graphs: advanced topics}.
\newblock MARCEL DEKKER, INC, 1998.

\bibitem{98dominate1}
T.~W. Haynes, S.~T. Hedetniemi, and P.~J. Slater.
\newblock {\em Fundamentals of domination in graphs}.
\newblock MARCEL DEKKER, INC, 1998.

\bibitem{63hoeffdingbound}
W.~Hoeffding.
\newblock Probability inequalities for sums of bounded random variables.
\newblock {\em Journal of the American Statistical Association},
  58(301):13--30, 1963.

\bibitem{12cacmsocialsearch}
D.~Horowitz and S.~D. Kamvar.
\newblock Searching the village: models and methods for social search.
\newblock {\em Commun. ACM}, 55(4):111--118, 2012.

\bibitem{03kddinfluence}
D.~Kempe, J.~M. Kleinberg, and {\'E}.~Tardos.
\newblock Maximizing the spread of influence through a social network.
\newblock In {\em KDD}, 2003.

\bibitem{07aaaiobserselection}
A.~Krause and C.~Guestrin.
\newblock Near-optimal observation selection using submodular functions.
\newblock In {\em AAAI}, pages 1650--1654, 2007.

\bibitem{08aaiprivacy}
A.~Krause and E.~Horvitz.
\newblock A utility-theoretic approach to privacy and personalization.
\newblock In {\em AAAI}, pages 1181--1188, 2008.

\bibitem{08jmlrsensorplacement}
A.~Krause, A.~P. Singh, and C.~Guestrin.
\newblock Near-optimal sensor placements in gaussian processes: Theory,
  efficient algorithms and empirical studies.
\newblock {\em Journal of Machine Learning Research}, 9:235--284, 2008.

\bibitem{03podcdisdominateset}
F.~Kuhn and R.~Wattenhofer.
\newblock Constant-time distributed dominating set approximation.
\newblock In {\em PODC}, 2003.

\bibitem{07socialbrowsing}
K.~Lerman.
\newblock Social browsing {\&} information filtering in social media.
\newblock {\em CoRR}, abs/0710.5697, 2007.

\bibitem{07icwsmsocialbrowsing}
K.~Lerman and L.~Jones.
\newblock Social browsing on flickr.
\newblock In {\em ICWSM}, 2007.

\bibitem{standforddata}
J.~Leskovec.
\newblock Standford network analysis project.
\newblock 2010.

\bibitem{07kddoutbreak}
J.~Leskovec, A.~Krause, C.~Guestrin, C.~Faloutsos, J.~M. VanBriesen, and N.~S.
  Glance.
\newblock Cost-effective outbreak detection in networks.
\newblock In {\em KDD}, 2007.

\bibitem{11icdmdivranklrh}
R.-H. Li and J.~X. Yu.
\newblock Scalable diversified ranking on large graphs.
\newblock In {\em ICDM}, 2011.

\bibitem{12tkdedivranklrh}
R.-H. Li and J.~X. Yu.
\newblock Scalable diversified ranking on large graphs.
\newblock In {\em IEEE Transactions on Knowledge and Data Engineering}, 2012.

\bibitem{10htldocsummarize}
H.~Lin and J.~Bilmes.
\newblock Multi-document summarization via budgeted maximization of submodular
  functions.
\newblock In {\em HLT-NAACL}, 2010.

\bibitem{11aclsubmodular}
H.~Lin and J.~Bilmes.
\newblock A class of submodular functions for document summarization.
\newblock In {\em ACL}, 2011.

\bibitem{07ccjdominate}
D.~Liu, X.~Jia, and I.~Stojmenovic.
\newblock Quorum and connected dominating sets based location service in
  wireless ad hoc, sensor and actuator networks.
\newblock {\em Computer Communications}, 30(18):3627--3643, 2007.

\bibitem{93randomwalkgraph}
L.~Lovasz.
\newblock Random walk on graphs: A survey.
\newblock {\em Combinatorics}, 2:1--46, 1993.

\bibitem{10icwsmsocialsearch}
M.~R. Morris, J.~Teevan, and K.~Panovich.
\newblock A comparison of information seeking using search engines and social
  networks.
\newblock In {\em ICWSM}, 2010.

\bibitem{78submodular}
G.~L. Nemhauser, L.~A. Wolsey, and M.~L. Fisher.
\newblock An analysis of approximations for maximizing submodular set
  functions-i.
\newblock {\em Mathematical Programming}, 14:265--294, 1978.

\bibitem{79connecteddomset}
E.~Sampathkumar and H.~Walikar.
\newblock The connected domination number of a graph.
\newblock {\em J. Math. Phys. Sci}, 13(6):607--613, 1979.

\bibitem{07uaihittingt}
P.~Sarkar and A.~W. Moore.
\newblock A tractable approach to finding closest truncated-commute-time
  neighbors in large graphs.
\newblock In {\em UAI}, 2007.

\bibitem{08icmlhittingt}
P.~Sarkar, A.~W. Moore, and A.~Prakash.
\newblock Fast incremental proximity search in large graphs.
\newblock In {\em ICML}, 2008.

\bibitem{10vldbsocialsearch}
X.~Si, E.~Y. Chang, Z.~Gy{\"o}ngyi, and M.~Sun.
\newblock Confucius and its intelligent disciples: Integrating social with
  search.
\newblock {\em PVLDB}, 3(2), 2010.

\bibitem{04tpdsdominateset}
I.~Stojmenovic, M.~Seddigh, and J.~D. Zunic.
\newblock Dominating sets and neighbor elimination-based broadcasting
  algorithms in wireless networks.
\newblock {\em IEEE Trans. Parallel Distrib. Syst.}, 13(1):14--25, 2002.

\bibitem{06tpdsextenddomset}
J.~Wu, M.~Cardei, F.~Dai, and S.~Yang.
\newblock Extended dominating set and its applications in ad hoc networks using
  cooperative communication.
\newblock {\em IEEE Trans. Parallel Distrib. Syst.}, 17(8):851--864, 2006.

\bibitem{08mobihocdomset}
Y.~Wu and Y.~Li.
\newblock Construction algorithms for k-connected m-dominating sets in wireless
  sensor networks.
\newblock In {\em MobiHoc}, 2008.

\end{thebibliography}
}

\end{document}